\documentclass{SBCbookchapter}
\usepackage[utf8]{inputenc}  
\usepackage[portuguese]{babel}
\usepackage{graphicx}
\usepackage{url}
\usepackage{color}
\usepackage{microtype}
\usepackage{comment}
\usepackage{multicol, blindtext}
\usepackage{multirow}
\usepackage[table,xcdraw]{xcolor}
\usepackage{units}
\usepackage{subfigure} 
\usepackage{hyperref}
\usepackage{booktabs}
\usepackage{array}
\usepackage[portuguese,linesnumbered,ruled]{algorithm2e}
\usepackage{times}
\usepackage{breakcites}

\author{Diego O. Rodrigues (IC-UNICAMP), Frances A. Santos (IC-UNICAMP), Geraldo P. Rocha Filho (CiC-UnB), Ademar T. Akabane (IC-UNICAMP), Raquel Cabral (UFAL), Roger Immich (IC-UNICAMP), Wellington L. Junior (IC-UNICAMP), Felipe D. Cunha (PUC-Minas), Daniel L. Guidoni (UFSJ), Thiago H. Silva (UTFPR), Denis Rosário (UFPA), Eduardo Cerqueira (UFPA), Antonio A. F. Loureiro (UFMG) e Leandro A. Villas (IC-UNICAMP)} 

\title{Computação Urbana da Teoria à Prática: \\ Fundamentos, Aplicações e Desafios}
 
\begin{document}
\maketitle 
\begin{abstract}
The growing of cities has resulted in innumerable technical and managerial challenges for public administrators such as energy consumption, pollution, urban mobility and even supervision of private and public spaces in an appropriate way. Urban Computing emerges as a promising paradigm to solve such challenges, through the extraction of knowledge, from a large amount of heterogeneous data existing in urban space. Moreover, Urban Computing correlates urban sensing, data management, and analysis to provide services that have the potential to improve the quality of life of the citizens of large urban centers. Consider this context, this chapter aims to present the fundamentals of Urban Computing and the steps necessary to develop an application in this area. To achieve this goal, the following questions will be investigated, namely: (i) What are the main research problems of Urban Computing?; (ii) What are the technological challenges for the implementation of services in Urban Computing?; (iii) What are the main methodologies used for the development of services in Urban Computing?; and (iv) What are the representative applications in this field?
\end{abstract}

\begin{resumo} 
\begin{otherlanguage}{portuguese}
A rápida urbanização das cidades vem resultando em inúmeros desafios técnicos e gerenciais para os gestores públicos tais como consumo de energia, poluição, mobilidade urbana e até mesmo gestão de espaços privados e públicos de forma apropriada. A Computação Urbana surge como um paradigma promissor para resolver tais desafios, por meio da extração de informação, a partir de uma grande quantidade de dados heterogêneos existentes no espaço urbano. Mais ainda, a Computação Urbana correlaciona o sensoriamento urbano, o gerenciamento de dados e sua análise para fornecer serviços que têm o potencial de melhorar a qualidade da vida dos habitantes de grandes centros urbanos. Com isso em mente, este minicurso tem como objetivo apresentar os fundamentos da Computação Urbana e os passos necessários para desenvolver uma aplicação nessa área. Para alcançar tal objetivo, serão investigadas as seguintes questões, a saber: (i) Quais são os principais problemas de pesquisa da Computação Urbana?; (ii) Quais são os desafios tecnológicos existentes para a implantação de serviços na Computação Urbana?; (iii) Quais são as principais metodologias utilizadas para o desenvolvimento de serviços na Computação Urbana?; e (iv) Quais são as aplicações representativas neste domínio?
\end{otherlanguage}
\end{resumo}

\pagestyle{plain}

\vspace{0.1cm}

\section{Introdução}
\label{sec:introducao}

O relatório da ONU, \cite{UN2018}, relata que $54\%$ da população mundial atualmente reside em áreas urbanas e a expectativa para 2050 é que esse número alcance $66\%$. 
Sabe-se que a rápida urbanização das cidades resulta em inúmeros desafios técnicos e gerenciais para os gestores públicos, tais como consumo de energia, poluição, mobilidade urbana e até mesmo gestão de espaços privados e públicos de forma apropriada~\cite{zheng2014urban,PRFILHO201854}.

Nesse contexto, a Computação Urbana surge como um paradigma promissor para superar desafios típicos dos grandes centros urbanos, principalmente, por meio da extração de informação de uma grande quantidade de dados heterogêneos existentes nas cidades~\cite{zheng2014urban}. 
Este campo de estudo visa aplicar soluções computacionais para melhorar os problemas enfrentados em grandes cidades. Para isso, a Computação Urbana utiliza diferentes tecnologias de sensoriamento remoto, técnicas de gerenciamento de dados e modelos analíticos, além de métodos de visualização para criar soluções que melhoram a qualidade de vida da população e dos sistemas de gerenciamento dos serviços oferecidos pelas cidades. Uma questão importante é que a Computação Urbana pode auxiliar a entender a natureza dos fenômenos urbanos ou até mesmo prevê-los~\cite{zheng2014urban,Silva2018}. 



\begin{figure*}[!htp]
\centering
  \includegraphics[width=1\textwidth]{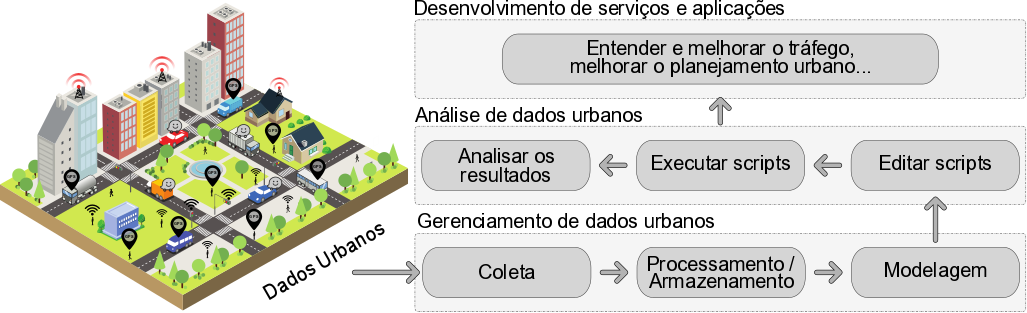}
\caption{Visão geral do arcabouço da computação urbana. [Inspirada em \cite{Silva2018}].}
\label{fig:arcabouco}
\end{figure*}

De uma maneira geral, pode-se apresentar a Computação Urbana composta por três macro atividades~\cite{zheng2014urban,Silva2018}: (i) gerenciamento dos dados urbanos; (ii) análise dos dados urbanos; e (iii) desenvolvimento de serviços e aplicações para cidades inteligentes, conforme exibido na Figura~\ref{fig:arcabouco}. O gerenciamento de dados urbanos consiste no processo de coleta de dados urbanos, no processamento desses dados e na modelagem dos mesmos. 
A análise dos dados urbanos é realizada por meio da implementação de algoritmos, bem como da interpretação de resultados. 
E, por fim, tem-se o desenvolvimento de serviços e aplicações a partir do conhecimento obtido. 

O minicurso tem como objetivo principal introduzir os fundamentos da Computação Urbana apresentando as ferramentas e os meios necessários para desenvolver uma aplicação nessa área. 
Os conceitos englobam uma visão geral dos trabalhos existentes, os desafios e as perspectivas futuras na área. A Computação Urbana é um tópico de pesquisa emergente, dessa forma existem problemas em aberto e desafios que não são triviais de serem resolvidos. Dessa forma, este minicurso também possui a finalidade de investigar as seguintes questões, as quais são retificadas por~\cite{zheng2014urban}: (i) Quais são os principais problemas de pesquisa da Computação Urbana?; (ii) Quais são os desafios tecnológicos existentes para a implantação de serviços na Computação Urbana?; (iii) Quais são as principais metodologias utilizadas para o desenvolvimento de serviços na computação urbana?; e (iv) Quais são as aplicações representativas neste domínio?

O restante deste minicurso está organizado da seguinte forma. A seção~\ref{sec:urbcom} apresenta os fundamentos teóricos e os desafios relacionados à Computação Urbana. A seção~\ref{sec:aplicacoes} apresenta as principais aplicações e serviços que podem ser desenvolvidas com base na Computação Urbana. A seção~\ref{sec:desafios} apresenta diversos tópicos de pesquisas relacionados com a Computação Urbana, os desafios e as questões de pesquisas em aberto. A seção~\ref{sec:estudodecaso} mostra uma perspectiva prática do desenvolvimento de um serviço com base em Computação Urbana. E por fim a seção~\ref{sec:conclusao} traz as conclusões a respeito do tema. 

\section{Computação Urbana} 
\label{sec:urbcom}


A demanda por aplicações relacionadas a computação urbana está em expansão. 
Conforme mencionado anteriormente, as Nações Unidas projetam que deverá haver um considerável aumento na população urbana nas próximas décadas~\cite{UN2018}.
Sendo assim, existe uma grande carência de técnicas que ofereçam uma infraestrutura adequada para as grandes cidades, tais como, serviços de transporte, habitação, água e energia.

Visando uma solução holística para o problema, a computação urbana faz uso de uma grande quantidade de fonte de dados, como de dispositivos da Internet das Coisas~(IoT)~\cite{IoT:2014,FILHO2019153}; dados de redes sociais baseadas em localização~(LBSN)~\cite{Silva2018,Zheng2011,Zheng2012,traynor2012location} e também dados estatísticos sobre cidades e sua população. 
Neste sentido, surge definição de computação em névoa, que oferece um diferencial importante em relação aos demais paradigmas, pois o processamento da informação está dispersa no espaço urbano, provendo assim uma rápida resposta e também a possibilidade de aplicações relacionadas ao contexto de cada região, como pode ser visto em~\cite{8291472,geraldo,geraldo2018sistema}. 
Por exemplo, sistemas de vigilância urbana podem receber suporte da névoa, permitindo automatizar tarefas e realizar a tomada de decisões em tempo real conforme a localização e contexto de cada região~\cite{chen2016dynamic}. 


De uma forma mais abrangente, a computação urbana busca os aspectos que permeiam os fenômenos urbanos, bem como fornecer estimativas sobre o futuro das cidades. Por este motivo, dentro da ciência da computação esse é um paradigma interdisciplinar, pois faz interseção com áreas que vão desde a de redes de computadores, redes de sensores e redes de veículos, interagindo também com sistemas distribuídos, inteligência artificial e redes sociais~\cite{Silva2018}. 

Conforme mencionado anteriormente, uma definição mais específica da computação urbana pode ser alcançada através de três macro atividades, nomeadamente (i) gerenciamento dos dados urbanos; (ii) análise dos dados urbanos; e (iii) desenvolvimento de serviços e aplicações para cidades inteligentes. 

A primeira etapa, (i) o gerenciamento dos dados urbanos, diz respeito a coleta e processamento das informações. Esta tem o objetivo de obter amostras de dados a partir de várias fontes de interesse de forma eficiente e contínua. A atividade de amostragem dos dados pode ser realizada a partir de fontes dinâmicas e heterogêneas, já que dados oriundos de apenas uma fonte pode não oferecer completude. Um dos desafios neste sentido é a criação de técnicas para reduzir os ruídos e erros durante a coleta de grandes fluxos de dados, tento em vista que a utilização de diferentes fontes pode resultar em dados conflitantes e/ou duplicados~\cite{Guo:TTT:2014}.  

Existem hoje, diversas fontes de dados publicamente disponíveis na Internet. Estas vão desde fontes com dados estatísticos sobre as cidades até dados das redes de sensoriamento participativo. 
Outra dificuldade nesta área é a diversidade dos formatos nos quais os dados estão sendo disponibilizados, pois não existe um padrão pré-definido para isto. Portanto, essa etapa também precisa se preocupar com processamento e formatação dos dados recolhidos. Além disto, buscando suprir a carência de dados em cidades que não tem programas de compartilhamento, é possível utilizar dados de redes sociais para obter informações similares aos dados estatísticos oficiais~\cite{silvaSocInfo2013,Silva2014toit}. 

Outra forma de obtenção de dados é através do sensoriamento participativo. Essas informações podem ser bastante úteis para o estudo de hábitos e rotinas de habitantes das cidades. Em especial, é possível ressaltar a importância destas informações para os ITS (\textit{Intelligent Transportation System}), uma vez que considerando os aspectos de mobilidade dos veículos e as suas trajetórias diárias, é possível extrair desse cenário diversos padrões de movimentação e também identificar pontos onde esta mobilidade está fluindo ou precisa ser revista. Com essas informações também é possível detalhar características culturais sobre as rotinas dos usuários, seus interesses e os pontos de maior visitação em uma cidade. 

A segunda etapa (ii), a análise dos dados urbanos, está posicionada após a coleta dos dados. Nessa etapa ocorre o processamento e a combinação de diferentes tipos de dados com o objetivo de extrair um conjunto de informações. Além disto, os dados brutos provenientes da etapa de sensoriamento requerem um processo de preparação dos dados, antes de prosseguir para a etapa de análise dos dados. Um dos grandes desafios nessa etapa é a elaboração e a implementação de um sistema capaz de analisar as correlações de um grande conjunto de dados, extrair informações e disponibilizá-los em tempo quase real. Assim como na primeira etapa, outro desafio que ainda persiste é em relação às fontes de dados heterogêneas, onde são necessários métodos e ferramentas para realizar a integração de diferente tipos de dados que podem ter sido obtidos das mais diversas fontes. 

A terceira etapa (iii), o desenvolvimento de serviços e aplicações para cidades inteligentes, refere-se aos procedimentos adotados para o consumo das informações. Para que esta atividade seja bem-sucedida é necessário o desenvolvimento de modelos descritivos e preditivos a partir das informações adquiridas. Os modelos descritivos podem ser utilizados para identificar relações entre as diversas variáveis coletadas, ou seja, ele serve para descrever o passado e auxiliar na tomada de decisões sobre o futuro. Esse modelo é muito útil para categorizar as informações e também identificar potenciais situações que podem ser ajustadas para prover benefícios aos serviços e/ou seus utilizadores. Uma das desvantagens dos modelos descritivos é que a intuição e a experiência acabam por te um peso excessivo na tomada de decisões. 

Devido a computação urbana analisar um grande fluxo de dados, estas características podem levar a resultados tendenciosos que não refletem de forma integral a realidade. Por isto, os modelos preditivos também deverão utilizados. Esses modelos conseguem identificar padrões ocultos e determinas as possíveis consequências resultantes desses padrões. Com este conhecimento em mãos, as decisões tomadas se tornarão mais assertivas. De uma forma geral, os modelos preditivos podem ser desenvolvidos através de métodos de aprendizado de máquinas, supervisionados ou não supervisionados. No primeiro, existe uma fase de treinamento do modelo, ou seja, são utilizados dados de entrada e de saída que descrevem a situação em particular que o método está sendo treinado para identificar. Por outro lado, nos modelos não supervisionados, somente os dados de entrada são introduzidos, sendo que o modelo deverá descobrir a relação entre estes dados para apresentar uma saída. 


O conteúdo de computação urbana é bastante abrangente. 
Neste capítulo vamos nos ater aos aspectos relevantes a este minicurso, a saber: computação em nuvem e névoa (Seção~\ref{subsec:nuvem}), gerenciamento e localidade dos dados~(Seção~\ref{subsec:localidade}), sensoriamento e aquisição de dados urbanos~(Seção~\ref{subsec:sensoriamento}) e por último, aspectos relacionados ao \textit{crowdsourcing} e \textit{crowdsensing}~(Seção~\ref{subsec:crowdsourcing}).

\subsection{Computação em Nuvem e Névoa} 
\label{subsec:nuvem}


A computação em Nuvem proporcionou grandes avanços nas formas de utilização das infraestruturas de Internet. Os seus centros de processamento de dados são, de uma maneira geral, amplas instalações em locais selecionados especificamente para esta finalidade, sendo que na sua maioria, encontram-se distribuídos através do globo terrestre. Cada centro de dados consolida um vasto número de equipamentos em um mesmo local físico, que pode oferecer serviços de forma isolada ou se comunicar com outros centros de dados para melhor atender o usuário final. A centralização destes recursos permite realizar uma economia financeira devido a escala de operação, viabilizando reduzir o custo de funcionado associado à sua instalação, manutenção e gerenciamento.



Uma funcionalidade importante da nuvem é a possibilidade de virtualização de equipamentos e serviços. Com isto, é possível prover uma maior elasticidade na alocação de recursos. Por conseguinte, os utilizadores podem requisitar somente a capacidade necessária para a realização das suas operações. Não existe a necessidade de efetuar um aprovisionamento maior do que o necessário, pois sempre é possível solicitar mais recursos caso haja necessidade. 


A convergência das redes de computadores com a nuvem auxiliou na resolução de diversos problemas, como por exemplo, provendo uma maior escalabilidade e disponibilidade, bem como aumentando a compatibilidade entre sistemas. Porém, ao mesmo tempo, introduziu novos desafios provenientes do distanciamento entre a produção e processamento dos dados, levando a uma sobrecarga do núcleo da rede e uma maior latência de comunicação~\cite{Curado2019}. 

Pelo fato de prover um serviço ubíquo, os seus usuários estão dispersos no mundo todo, por conseguinte, muitos destes se encontram geograficamente distantes dos provedores da nuvem. Esta característica pode não ter grande impacto para o usuário convencional, que utiliza apenas os recursos mais simples disponíveis, como por exemplo, armazenamento de arquivos, redes sociais e navegação na Internet. Por outro lado, isto pode representar um grande desafio para dispositivos restritos, que possuem pouca capacidade de memória, processamento e comunicação, características estas de uma considerável parte dos dispositivos pertencentes a IoT (Internet of Things) que são amplamente usados em cidades urbanas e inteligentes. 

Dependendo da aplicação, mesmo em dispositivos de maior capacidade, esta distância pode se tornar um impedimento para atingir plenamente os requisitos dos serviços disponibilizados, como por exemplo, em previsão de congestionamento e gerenciamento de energia. Desta forma, é necessário trazer ao menos uma parcela do poder computacional mais próximo dos usuários e/ou serviços. Por isto, a computação em névoa vem ganhando força~\cite{Byers2017}.  A proposta básica deste paradigma é trazer a elasticidade de recursos da nuvem para perto dos usuários/objetos/serviços.

\begin{figure*}[!htp]
\centering
  \includegraphics[width=11cm]{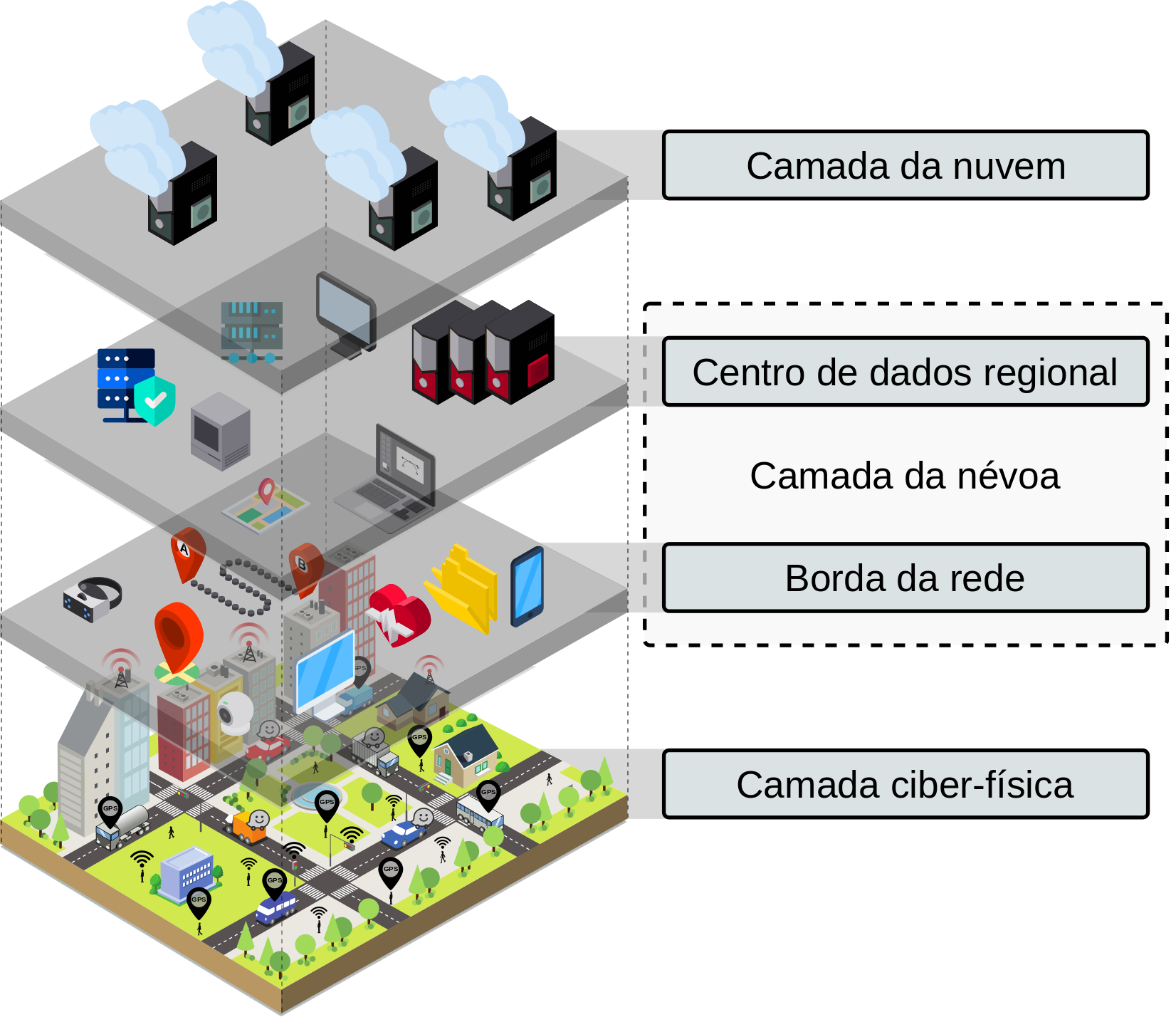}
\caption{Arquitetura de computação urbana com múltiplas camadas de névoa e nuvem}
\label{fig:arquiteturaNevoaCompUrbana}
\end{figure*}

A utilização conjunta dos paradigmas de nuvem e névoa pode ser organizada em uma arquitetura de camadas, conforme definido pelo consórcio Openfog~\cite{openfog_2017} e ilustrado na Figura~\ref{fig:arquiteturaNevoaCompUrbana}. Um dos modelos aceitos hoje em dia é formado por três camadas, que são: a camada ciber-física, camada da névoa e a camada da nuvem. A camada inferior é composta pelos dispositivos físicos, sejam eles móveis ou estáticos. Com o aumento do poder computacional destes equipamentos e o aprimoramento das tecnologias de comunicação, estes são capazes de executar processamentos relativamente complexos com uma boa performance~\cite{Taleb2017}. 

Estes aparelhos, como por exemplo \textit{smartphones}, \textit{smartwatches}, consoles de jogos, óculos inteligentes, entre outros, em conjunto com os demais servidores e equipamentos, da borda da rede, localizados próximos aos usuários, podem ser utilizados para aliviar o ambiente da nuvem e ainda prover menor latência na comunicação. 

A camada intermediária (névoa), é composta por múltiplos níveis, com o objetivo de fornecer uma maior flexibilidade na composição da sua infraestrutura. Sendo assim, com o estabelecimento desta hierarquia de níveis é possível oferecer um equilíbrio em algumas das características mais desejáveis, como a capacidade computacional, tempo de acesso e responsabilidade organizacional. De uma forma geral, os níveis mais próximos da borda da rede tendem a possuir uma menor responsabilidade organizacional na infraestrutura e, portanto, uma menor cobertura geográfica. Por outro lado, níveis mais próximos da nuvem possuem uma maior cobertura geográfica e capacidade computacional, mas tendem a apresentar tempos de acesso mais elevados. A camada superior é composta pelos servidores de nuvem, os quais podem estar localizados em nuvens públicas, privadas, ou uma combinação de ambas. 

É válido mencionar que um dos maiores desafios em uma rede centrada na computação em névoa é como alcançar o melhor desempenho possível considerando o poder de processamento e comunicação de cada equipamento, em conjunto com o processamento e comunicação oferecidos pela névoa. Isto deve-se ao fato de que são esperados dispositivos com características heterogêneas nestas redes. A computação em névoa precisa levar em consideração a heterogeneidade tanto das tecnologias de rede como dos dispositivos, permitindo que estes coexistam e operem em conjunto para prover níveis satisfatórios de qualidade. Por isto, um amplo conjunto de parâmetros devem ser contemplados, como por exemplo, dispositivos móveis com capacidade de bateria limitadas, diferentes formas e capacidades de comunicação e também distintas capacidades de processamento~\cite{BITTENCOURT2018}. 

Neste sentido, um aspecto importante é a Qualidade de Experiência~(QoE) dos usuários finais. Este conjunto de métricas busca definir de uma forma quantitativa a aceitação de um serviço/aplicação, considerando desde a origem das informações até como elas são apresentadas na sua forma final~\cite{Immich2018,7179389}. Por isto, para uma solução ser considerada adequada para a computação em névoa, esta também deve levar em consideração a expectativa de QoE dos usuários finais e o seu impacto no sistema como um todo. 

Além de desafogar o ambiente de nuvem e oferecer comunicação de baixa latência, a computação em névoa também pode prover serviços cientes de sua geolocalização. 
Tendo em vista que o processamento das informações será realizado na borda da rede, ou seja, em uma localização geográfica muito próxima ao usuário, diversos detalhes sobre a rede de acesso 
podem ser utilizadas, como a qualidade de sinal da borda até o usuário e também a carga atual da célula em que o usuário se encontra. 
Todas estas informações podem se apresentar como um grande diferencial para a maioria das aplicações de computação urbana.

\subsection{Gerenciamento e Localidade dos dados} 
\label{subsec:localidade}

Conforme mencionado anteriormente, a produção e consumo de dados está atingindo números recordes e continua com ritmos de crescimento sem precedentes. 
Como resultado desta tendência, o gerenciamento e localidade dos dados vêm ganhando muita atenção ultimamente. De uma forma simplificada, estes conceitos se referem à capacidade de organizar e manter processos relacionados a aquisição, processamento, distribuição e armazenamento de dados, bem como a proteção e validação de informações. Esses conceitos foram pesquisados no passado em diferentes contextos, como por exemplo, em \textit{clusters} computacionais e também nas áreas de computação paralela e distribuída. No entanto, até recentemente estes princípios ainda não haviam sido aplicados em centro de dados geograficamente distribuídos, como os presentes na computação em nuvem e névoa~\cite{Xu2013,Hung2015,Heintz2016}. 

O gerenciamento de dados faz alusão ao planejamento e implantação de uma série de políticas e procedimentos para realizar a gestão completa, precisa e integral do ciclo de vida dos dados. De uma forma geral, esta atividade pode ser dividida em duas estratégias distintas, as quais definem (i) o posicionamento e (ii) a forma de acesso~\cite{Sakr2011}. A estratégia de posicionamento busca especificar onde estas informações serão armazenadas e como serão disseminadas. Esse procedimento, por exemplo, é responsável por determinar quantas cópias dos dados devem ser realizadas e também quais são os melhores locais/dispositivos para armazenar essas informações. 
A segunda estratégia define como deverão ser tratadas pelo sistema as operações de leitura e escrita. Para esta finalidade, é necessário considerar conjuntamente os procedimentos que deverão manter a consistência dos dados entre cópias distribuídas. 
Outra responsabilidade é também a de definir as formas de acesso as informações replicadas. 

Assim como o gerenciamento de dados, o conceito de localidade é igualmente importante. Este pode ser definido como a habilidade de mover ou alocar a capacidade computacional necessária para o processamento de informações nos locais próximos de onde estão sendo criados e/ou adquiridos estes dados~\cite{Yang2017}. Desta forma, é possível evitar a transferência indiscriminada de uma grande quantidade de dados brutos diretamente até recursos computacionais centralizados, como por exemplo, servidores da nuvem.
Esta estratégia baseia-se no fato de que movendo e executando os aplicativos computacionais para próximo dos dados sobre os quais eles operam é possível realizar uma computação mais eficiente e ainda prover a economia de recursos de rede. 
Com esta estrutura, consequentemente, elimina-se a necessidade de transmitir grandes quantidades de dados não processados até o aplicativo/serviço. 
É importante ressaltar que está estratégia é válida para serviços que necessitam de uma grande quantidade de dados para operar ou, os quais precisam fornecer uma resposta ultrarrápida as solicitações dos usuários. Por este motivo, os fundamentos da localidade de dados se encaixam com uma das principais premissas da computação em névoa, que é ter um sistema descentralizado com recursos próximos aos usuários finais~\cite{Wen2017}. 

Nesta perspectiva, pesquisas confirmam que o armazenamento de dados nas bordas da rede, utilizando a computação em névoa, pode melhorar o tempo de resposta e reduzir o tráfego de rede~\cite{Confais2017}. Além das questões de desempenho, a localidade dos dados pode aprimorar mecanismos relacionados à segurança e privacidade das informações~\cite{Bellavista2017}. Operando localmente, é possível ter um conhecimento preciso da estrutura da rede e implementar mais facilmente recursos de autenticação e autorização. Outra vantagem de ter os dados próximos de onde estes serão processados é o aperfeiçoamento das questões de privacidade~\cite{Vaquero:2014:FYW:2677046.2677052}, pois as informações não precisam ser transferidas através da rede, ou seja, existe um maior controle de acesso sobre as mesmas. 


Outra vantagem da localidade dos dados é a possibilidade de fazer o descarregamento de operações, tais como processamento ou armazenamento. Mesmo com o avanço nas tecnologias móveis, em algumas situações ainda é vantajoso transferir o processamento do próprio dispositivo para a borda da rede visando uma economia de bateria e a redução no tempo de processamento~\cite{Gama2018}. 
A computação em névoa auxilia nesta atividade provendo camadas de recursos entre os usuários e a nuvem~\cite{yi2015fog,shi2016edge}. 
É importante ressaltar que com a dispersão dos recursos através de toda a hierarquia de rede, são necessários novos mecanismos de gerenciamento de recursos, bem como métodos de migração para garantir que as aplicações e dados descarregados estão sempre o mais próximo possível dos usuários~\cite{Taleb2017b}.

Na computação em nuvem, a migração de máquinas virtuais, serviços e dados pode ser utilizada para realizar o balanceamento de carga de centro de dados e também para consolidar estas operações em uma quantidade menor de servidores. 
De mesma forma, na computação em névoa, a migração destes serviços/aplicações pode ser utilizada para replicar e/ou mover tanto dados como poder de computação para próximo dos usuários que necessitam ou poderão necessitar em um futuro iminente. 
Através da computação urbana, por exemplo, pode ser identificado os principais locais onde os usuários costumam realizar acessos e consumir informações.
Tendo em vista que estes padrões de comportamento da mobilidade humana podem ser previsto de uma forma consistente~\cite{song2010limits}, é possível carregar antecipadamente os dados e serviços nesses locais, reduzindo consideravelmente o tempo de acesso aos mesmos~\cite{7424534}.

\subsection{Sensoriamento e aquisição de dados urbanos} 
\label{subsec:sensoriamento}

A extração de conhecimento a partir dos dados consiste, basicamente, na modelagem e análise dos dados definindo uma semântica para a tomada de decisão para um determinado serviço~\cite{Barnaghi2012}, por exemplo, a definição de rotas mais rápidas e seguras. 
A aquisição e processamento de dados tem sido um dos grandes desafios em Computação Urbana, pois os dados são heterogêneos, provem de diversas fontes, possuem diferentes formatos (tabelas, mapas, grafos, etc) e são de diferentes tipos (dados sobre o tráfego, qualidade do ar, dados sociais e dados geográficos, etc). 
De acordo com a Figura~\ref{fig:arcabouco} as etapas necessárias para o tratamento dos dados até que eles estejam disponíveis para a aplicação são a coleta dos dados, processamento, armazenamento e modelagem. Estas etapas são detalhadas a seguir. 

A \textit{coleta de dados} é a primeira etapa a ser realizada.
De acordo com~\cite{Silva2018}, os dados usados nessa área são provenientes de quatro fontes: 
\begin{itemize}
\item Dados de sensores: atualmente, as redes de sensores são muito utilizados para coleta de dados em ambientes urbanos.
Elas fornecem dados que são obtidos através da instalação de sensores dedicados a aplicações específicas, por exemplo, sensores para monitoramento da qualidade do ar em diversos pontos da cidade ou sensores para o monitoramento de níveis de ruídos. Atualmente, é muito comum o uso de veículos conectados e dispositivos móveis para a obtenção de dados.  
Nesse caso, deve-se considerar o custo existente para a construção das redes de sensoriamento e que os veículos devem estar previamente equipados.  
\item Dados da infraestrutura disponível nas cidades: atualmente as cidades estão construindo infraestruturas para diversos propósitos, por exemplo a rede de telefonia celular, redes sem fio ou ainda sistemas de transporte público. Muitas vezes, essa infraestrutura também pode ser utilizada para coletar dados. 
\item Dados estatísticos: são dados referentes a estudos estatísticos sobre uma população, tais como dados demográficos, econômicos e sociais relativos a um momento determinado ou em certos períodos. O trabalho de~\cite{DeSouza2018} implementa uma ferramenta para sugestão de rotas usando, além de dados sobre a condição de tráfego, dados relacionados a crimes na cidade de São Paulo. Os dados foram obtidos do banco de dados oficial do departamento de policia. 
Vale ressaltar, que muitas vezes não existe a disponibilidade desses dados e quando disponíveis eles possuem diversos formatos ou são incompletos. Assim, deve existir um mecanismo para obtenção e tratamento desses dados antes de sua utilização.
\item Dados de redes sociais: no ambiente de computação urbana os dados das mídias sociais, tais como Twitter, \textit{Instagram}, \textit{Waze} e \textit{Foursquare} são usados para modelar a mobilidade das pessoas nas áreas urbanas e são usados em diversas aplicações, tais como na identificação de problemas no trânsito causados por acidentes, protestos, desastres, etc \cite{silvaSocInfo2013,Ribeiro2014,santosWI2018,RODRIGUES2018111}.
Esses dados possuem diversos formatos, tais como textos, fotos e vídeos. 
Aqui está incluído, o sensoriamento participativo, que é o processo pelo qual indivíduos e comunidades usam celulares e serviços de nuvem para compartilhar voluntariamente informações~\cite{Estrin2010}. Vale destacar que a obtenção desse tipo de dado pode ser mais fácil do que os três anteriores, dada a grande quantidade desses dados disponíveis na WEB. Esses dados podem ser obtidos por meio de APIs disponibilizadas pelas ferramentas; através de \textit{Web crawler}, que são programas que analisam páginas Web em busca de dados relevantes;  ou desenvolvendo aplicações em plataformas já existentes, por exemplo, algumas ferramentas, como Facebook e Instagram, permitem a criação de aplicativos dentro de suas plataformas.
\end{itemize}

Dessa forma, existem diversas maneiras para coletar dados no ambiente de computação urbana, nesse minicurso serão usados dados de redes sociais para exemplificar o desenvolvimento de uma aplicação no ambiente de computação urbana. 

Uma vez que esses dados são coletados, deve-se utilizar técnicas de \textit{processamento e armazenamento}, já que essa etapa envolve a manipulação de um grande volume de dados.  
Em alguns casos, os dados coletados são suficientes para revelar um padrão interessante, não sendo necessário um processamento complexo desses dados. No entanto, na maioria das vezes esses dados são provenientes de diversas fontes e possuem formatos diferentes, dessa forma são necessários técnicas específicas para formatação desses dados. Por exemplo, no problema de geração de rotas em ambientes urbanos podemos considerar as informações de trânsito disponibilizadas pela infraestrutura da cidade e informações a respeito de áreas de perigo.
Dependendo da aplicação, será necessário a comparação dos dados atuais com tendências passadas, de modo que o armazenamento e o gerenciamento robusto e de longo prazo desses dados são um requisito central. 
De uma forma geral, nessa etapa são desenvolvidos \textit{scripts} para a formatação e organização dos dados. 
A organização dos dados nos leva a entender melhor esses dados e definir quais os modelos e análise são apropriados para serem aplicados~\cite{Silva2018}. 
Na Seção~\ref{sec:estudodecaso} será mostrado um exemplo prático do processamento de um tipo de dados específico. 

O armazenamento envolve sistemas de arquivo e ferramentas para processamento distribuído de grandes volumes de dados e deve ser: escalável, distribuído, seguro, tolerante a falhas e consistente. No caso da computação em névoa, os dados são movidos para o melhor lugar para processamento, isto é, mais perto do limite da rede. 
Portanto é uma arquitetura de computação descentralizada onde dados, cálculos, comunicações, armazenamentos, medições, aplicações e gerenciamentos são distribuídos entre a fonte de dados e a nuvem. Dessa forma, tem-se uma redução de dados que necessitam ser transportados para análise, processamento ou armazenamento e maior velocidade no acesso aos dados pelo usuário final. 

Algumas técnicas são utilizadas para \textit{modelagem} dos dados. Os dados aqui possuem propriedades tanto espaciais, como temporais. 
Por exemplo, a localização de estabelecimentos é um dado espacial; dados meteorológicos, vídeos de vigilância e consumo de energia são dados temporais. 
Já os dados como como fluxos de tráfego e mobilidade humana possuem propriedades espaço-temporais simultaneamente~\cite{zheng2014urban}.
Nessa etapa os dados devem ser organizados em alguma estrutura que incorpora simultaneamente o dado e as informações espaço-temporais para suportar uma análise de dados eficiente. 
Além disso, um sistema de computação urbana geralmente precisa aproveitar uma variedade de dados heterogêneos e responder rapidamente às consultas dos usuários, por exemplo, as condições de tráfego em uma determinada área. 
Levando em conta as propriedades espaço-temporais desses dados, temos que encontrar modelos que consigam capturar a dinâmica dos dados. 
Em computação urbana, é comum a utilização de grafos para representar a propriedade espacial. 
Por exemplo, em uma rede veicular mapeada como um grafo, cada veículo representa um nó do grafo e as ligações são estabelecidas com o nó se dois veículos conseguem se comunicar.  
O conteúdo desses dados pode variar muito ao longo do tempo, assim um modelo baseado em grafos estáticos não é suficiente para capturar essa dinamicidade, dessa forma a teoria de Redes Complexas tem sido utilizada para representar diversos problemas nesse contexto. 
Essa teoria provê modelos (redes de mundo pequeno, redes livres de escala) e métricas (p.e., centralidade de nós, coeficiente de agrupamento) que auxiliam no entendimento da dinâmica do problema~\cite{Naboulsi2013,Moura2018}.

A seguir pode ser realizada a análise de dados para se identificar anomalias, tais como os locais onde a mobilidade das pessoas difere significativamente de seus padrões de origem. 
Dessa forma, vamos caracterizar os dados coletados, a fim de entender suas limitações e sua utilidade. 
Geralmente essa etapa é composta pelas etapas de extração e análise da informação obtida a partir dos dados brutos e validação de resultados. 
A extração de conhecimento pode explorar diferentes abordagens, dependendo do problema que estamos tentando resolver. 
Um procedimento típico realizado é a investigação preliminar dos dados para entender suas propriedades, auxiliando na escolha de técnicas de análise. 

As técnicas de visualização, como histogramas e diagramas de dispersão, medidas estatísticas, como média e desvio padrão de um conjunto de valores, são métodos comuns usados para explorar propriedades de dados nesta investigação preliminar~\cite{Silva2018}.

Algumas tecnologias e ferramentas devem ser utilizadas nessa etapa para a análise dos dados. Técnicas como, algoritmos de agrupamento, regressões, modelos probabilísticos e análise de sentimento são muito utilizadas nesse contexto. 
Em relação as ferramentas para análise, as linguagens Python, Matlab/Octave e R são robustas, eficientes e de fácil compreensão para o desenvolvimento dos algoritmos.
 
\subsection{Crowdsourcing e Crowdsensing} 
\label{subsec:crowdsourcing}



Conforme mencionado anteriormente, uma das principais funções da computação urbana é referente ao processo de aquisição, integração e análise de grande volume de dados. 
Na sua essência, estes dados costumam ser heterogêneos, ou seja, não seguem nenhum padrão definido, e gerados por fontes em diferentes espaços urbanos. 
Além das tecnologias, modelos analíticos e métodos aplicados na Computação Urbana, dois importantes paradigmas podem ser também aplicados: \textit{crowdsourcing} e \textit{crowdsensing}. A seguir uma breve explicação de cada um deles.

\begin{figure}[htp]
\centering
\includegraphics[width=1\textwidth]{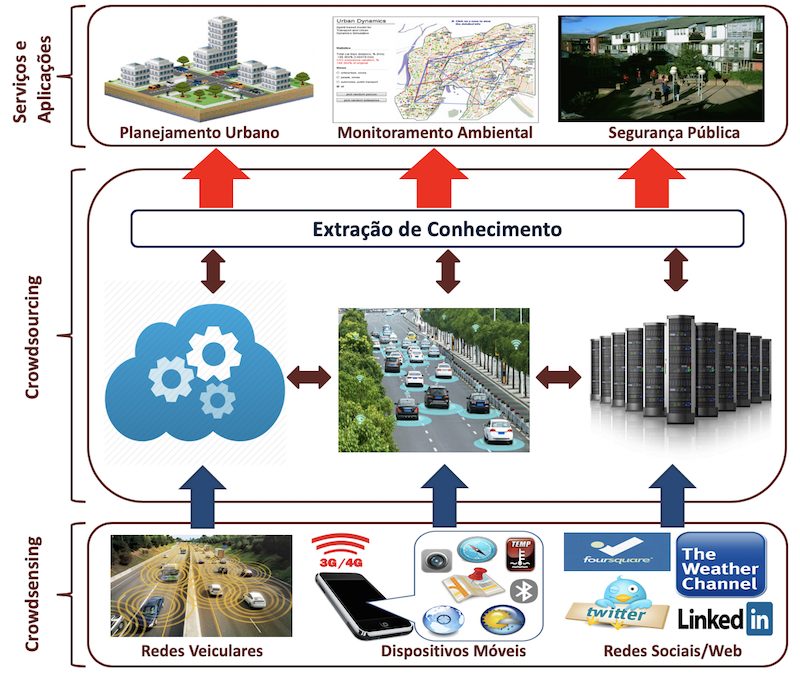}
\caption{Uma arquitetura ilustrativa de \textit{crowdsourcing} e \textit{Crowdsensing}.}
\label{fig:crowdsourcing}
\end{figure}

\textit{Crowdsensing} móvel é um paradigma que utiliza o conceito de computação ubíqua no sensoriamento do ambiente, além de compartilhar os dados sensoriados~\cite{wang2018city,akabane2018imob,akabane2018trusted}. Na primeira camada da arquitetura da Figura \ref{fig:crowdsourcing} apresenta um exemplo de \textit{crowdsensing}, onde diferentes dispositivos são utilizados para extrair coletar dados brutos. Além disso, por meio da agregação de tais dados, pode-se criar uma consciência local que pode auxiliar aplicações de larga escala, por exemplo, monitoramento da poluição do ar e alerta de congestionamento de tráfego veicular. Sabe-se que nas \textit{Vehicle Social Networks} - VSNs, os veículos são equipados com tecnologias de computação e de comunicação sem fio juntamente com sensores inteligentes, assim possibilitando o surgimento do paradigma de \textit{vehicle crowdsensing} - VCS~\cite{akabane2018imob}. Tal paradigma possibilita o monitoramento de fenômenos dinâmicos e em grande escala. A motivação do uso do \textit{crowdsensing} móvel está no fato de que os participantes possam mapear fenômenos de interesse comum. Por exemplo, os participantes de uma VSNs podem conjuntamente monitorar as condições do tráfego veicular, e também, aprimorar a mobilidade urbana por meio do compartilhamento dos dados sensoriados. Ao fazer isso, aplicações de VSNs podem agregar os dados locais sensoriados e gerar conhecimento sobre as condições de tráfego em tempo real, e tal conhecimento auxiliará o gerenciamento da mobilidade urbana. Outro exemplo \cite{issarny2018service} é o monitoramento da exposição à poluição em ambiente urbano.

O \textit{crowdsourcing} é um paradigma de cooperação, no qual um grande grupo de usuários (sendo que cada usuário resolve pequenas sub-tarefas de um trabalho maior) trata de problemas complexos que podem ser resolvidos, por meio da participação de muitos usuários, de maneira eficiente, por exemplo, gerenciamento inteligente de tráfego veicular \cite{akabane2018imob}). Na camada do meio da arquitetura da Figura \ref{fig:crowdsourcing} apresenta um exemplo de \textit{crowdsourcing}, onde diferentes tecnologias são empregadas para extrair conhecimento dos dados sensoriados.
Outra ideia muito aplicada é a utilização de sensores móveis no contexto de \textit{crowdsourcing}, denominado de \textit{Mobile Crowdsourcing} (MCS). O MCS difere das redes de sensores tradicionais, pois estes envolvem pessoas em movimento que estão processando dados ao seu redor de diferentes locais. Neste caso, as pessoas não estão apenas transportando sensores integrados aos seus dispositivos móveis (\textit{smartphones}), mas também são capazes de fornecer informações manualmente, como fotos e vídeos. As vantagens do MCS em relação a outros tipos de redes de sensores incluem: (i) alta capacidade de computação dos dispositivos móveis para pré-processamento de dados; (ii) conectividade com nuvem ou névoa; e (iii) usuários podem fornecer informações do ambiente que não são coletadas pelos sensores. Com isso surgem sistemas que identificam anomalias urbanas baseado em \textit{crowdsourcing}. Tais sistemas permitem a comunicação pervasiva e em tempo real das anomalias que ocorrem na cidade (por exemplo, ruído, uso ilegal de instalações públicas, mau funcionamento da infraestrutura urbana).

Vale salientar que vários sistemas e aplicativos de \textit{crowdsourcing} e \textit{crowdsensing} desenvolvidos no paradigma da Computação Urbana são dedicados a permitir que os cidadãos colaborem na melhoria da qualidade de vida no ambiente urbano.

\section{Aplicações}
\label{sec:aplicacoes}

Esta seção apresenta algumas das principais aplicações e serviços no contexto de computa-ção urbana em diferentes domínios. O objetivo é mostrar que a integração e análise de dados obtidos de diferentes fontes em espaços urbanos têm um papel importante em prol do desenvolvimento das cidades inteligentes. Com isso, a Seção \ref{secSTI} apresenta os sistemas de transportes inteligentes. A Seção \ref{secLBSNs} discute aplicações e serviços sociais baseadas em localização. Por fim, a Seção \ref{secSegPub} apresenta as aplicações no contexto de segurança pública.

\subsection{Sistemas de Transportes Inteligentes}
\label{secSTI}

O setor de transporte é um dos principais propulsores para desenvolvimento econômico de uma nação. Tanto cidadãos quanto empresas dependem desse setor no dia a dia. Por exemplo, as empresas dependem dos meios de transporte para obter os insumos, além de distribuir seus produtos até os consumidores. Já os cidadãos necessitam de tais meios para o deslocamento diário da casa para o trabalho e vice-versa.

Com os avanços das tecnologias de comunicação sem fio e da computação móvel permitiram o surgimento dos sistemas de transporte inteligentes (ITS). Tais sistemas têm como objetivo principal melhorar a mobilidade urbana, além de aumentar a segurança no trânsito. 
Aplicações desenvolvidas para ITS, geralmente, visam auxiliar os condutores e os passageiros durante seus deslocamentos, com intuito de reduzir acidentes (\textit{aplicação de segurança}) e melhorar o gerenciamento de tráfego veicular (\textit{aplicação de eficiência de tráfego})~\cite{meneguette2016increasing,meneguette2016solution}. Além dessas, há outro tipo, \textit{aplicação de entretenimento e conforto}, como o próprio nome sugere, esta tem o objetivo de tornar a viagem mais prazerosa e tranquila. Para mais detalhes de cada aplicação mencionada anteriormente, consulte \cite{cunha2017sistemas}. 

\begin{figure}[!htbp]
\centering
\includegraphics[width=1\textwidth]{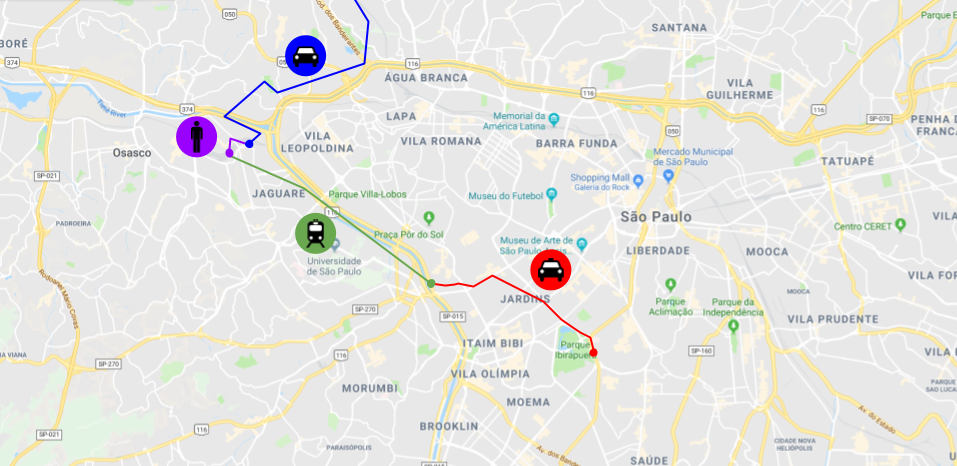}
\caption{Um exemplo ilustrativo de transporte multimodal.} 
\label{fig:multimodal}
\end{figure}

Outra aplicação de ITS é o \textit{transporte multimodal} (TM), foco do estudo de caso deste minicurso (apresentado na Seção 1.5). A Figura \ref{fig:multimodal} apresenta um exemplo ilustrativo de TM, no qual utiliza diversas modalidades de transporte (por exemplo: táxi, ônibus, trem ou metrô) para transportar cargas ou passageiros entre os pontos de origem e de destino \cite{zografos2008algorithms,steadieseifi2014multimodal}. Por consequência, o TM oferece um serviço que permite o transporte mais eficiente, confiável, flexível e sustentável. Sabe-se que, muitos cidadãos dos grandes centros urbanos necessitam utilizar diferentes tipos de transporte público multimodal nos seus deslocamentos diários. Logo, encontrar o caminho com menor custo, entre dois pontos, levando em consideração mais de um modo de transporte é o principal desafio do Problema de Transporte Multimodal.

Alguns trabalhos da literatura lidam com esse tipo de problema. Por exemplo, no trabalho de \cite{duque2015exact}, foi aplicado um algoritmo exato para a seleção dos caminhos. Para isto, duas variáveis conflitantes, custo e tempo, são aplicadas na solução. O algoritmo proposto foi inspirado no ``Algoritmo de Pulso'' \cite{birari2005mitigating}, ou seja, envia um pulso na rede e mensura o caminho realizado por ele, em relação aos pesos dos arcos até ao destino. Utilizando as mesmas variáveis conflitantes, no trabalho de \cite{sedeno2015dijkstra} foi aplicado um algoritmo de caminho mínimo e a solução é classificada por meio do conceito da não dominância de uma das variáveis. 

No trabalho de \cite{idri2017new}, os autores propõem um algoritmo ``orientado ao alvo''. Tal algoritmo aplica a busca pela proximidade dos nós em relação ao destino, ou seja, nessa etapa constrói-se o caminho virtual. O cálculo da proximidade é feito a partir da distância Euclidiana entre os pontos de origem e de destino. Além disso, durante o cálculo é aplicado o conceito de desigualdade triangular para priorizar a escolha dos nós localizados mais próximo do destino. Outra abordagem concebida pelos autores \cite{dib2015memetic} é combinar duas meta-heurísticas (Algoritmos Genético e de Busca Local), com intuito de simular o Algoritmo Memético \cite{krasnogor2005tutorial}. Tal algoritmo, faz uma analogia à memória humana, nas quais são memorizadas/armazenadas as informações mais importantes do processo, para que no final seja possível recuperar tais informações.

\subsection{Aplicações sociais baseadas em localização}
\label{secLBSNs} 

Atualmente existem diferentes tipos de redes sociais na Internet com diversos propósitos. Nesta seção, nosso foco são as redes sociais baseadas em localização (LBSN), um tipo especial de rede social em que informações geo-localizadas representam um papel fundamental no sistema. O estudo de LBSNs é interessante, pois elas geram um tipo específico de dados digitais que oferecem resoluções geográficas e temporais sem precedentes. Com isso, os dados provenientes desses sistemas podem ser úteis para entender diversos fenômenos urbanos em larga escala e com um custo, potencialmente, mais baixo \cite{Silva2018}. O Foursquare, Waze e Instagram são exemplos clássicos de LBSNs, no entanto existem vários outros exemplos um pouco menos conhecidos, como o Untappd\footnote{https://untappd.com.} que é uma rede especializada cervejas. 

De acordo com~\cite{Zheng2011}  uma rede social baseada em localização consiste em uma nova estrutura social composta por indivíduos ligados pela interdependência derivada de suas localizações físicas, bem como o seu conteúdo (como fotos, vídeos e textos) geo-localizados. 
Essa interdependência existe, se por exemplo, os indivíduos têm históricos semelhantes de localização e interesses em comum (i.e., comportamentos e atividades). 

As LBSNs contam com dados geográficos coletados em tempo real de forma voluntária e com o auxílio de dispositivos móveis dos indivíduos. 
Esse é um dos fatores que favorecem a alta escalabilidade dessas redes \cite{silvaUserUrbSensing}. Esses dados podem ser úteis para prover, por exemplo, entretenimento e informações de tipos variados (e.g., para tomar melhores decisões no trânsito). 

Uma etapa importante para o desenvolvimento de diversas aplicações é a estimativa de semelhança entre usuários utilizando os dados provenientes de LBSNs, nesse caso, é importante descrever o comportamento e interesses dos usuários. 
Existem várias maneiras para estimar a semelhança entre usuários. De uma forma geral, uma boa estimativa de semelhança depende da disponibilidade de boas informações descritivas, tais como informações de visitas e popularidade de locais, bem como a granularidade geoespacial das informações~\cite{zheng2014urban}. 
O comportamento e interesses de um indivíduo podem ser descritos, por exemplo, pelo seu histórico de \textit{check-ins}\footnote{ato de disponibilizar a sua localização atual para amigos na rede.} na cidade. Assim, pessoas que compartilham informações semelhantes de localização tendem a ter interesses e comportamentos comuns, a partir dessas informações pode-se detectar comunidades. Essa informação é útil, por exemplo, na realização de propagandas direcionadas, ou ainda criar sistemas de recomendação para diversos tipos de conteúdo. 

\begin{figure}[!htbp]
\centering
\includegraphics[width=1\textwidth]{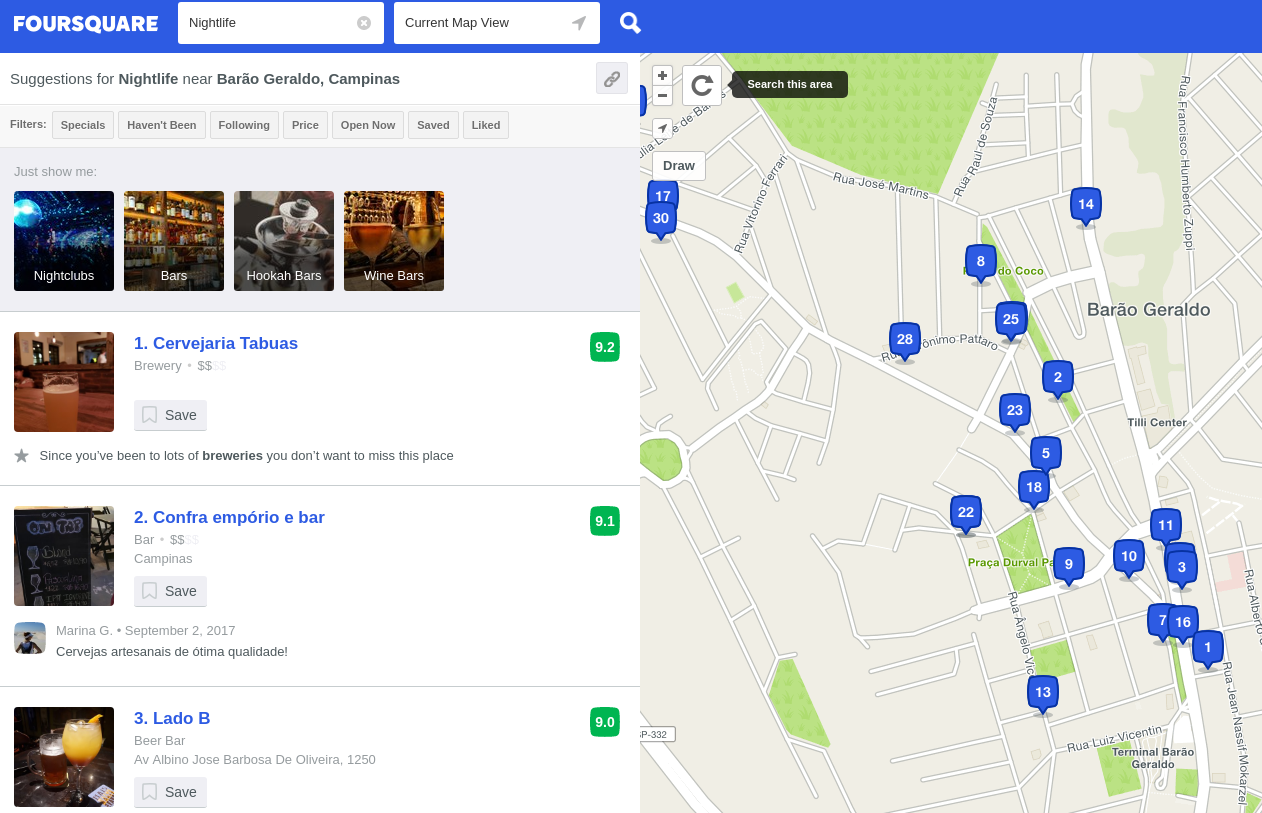}
\caption{Um exemplo ilustrativo de recomendação de locais fornecida pelo Foursquare. }
\label{figFoursquare}
\end{figure}

Um recomendador de lugares é um exemplo específico de aplicação de recomendação que consiste em encontrar e recomendar os locais mais interessantes em uma cidade. 
Vários fatores precisam ser levados em consideração para oferecer uma boa recomendação de locais. O nível de interesse de um local não depende exclusivamente do número de pessoas que visitaram o mesmo. Por exemplo, o local mais frequentemente visitado em uma cidade pode ser uma estação rodoviária ou aeroporto, mas esse tipo de local tende a não ser uma localização interessante de ser recomendada, por exemplo, a um turista que está visitando uma cidade. Com isso, além da popularidade, é importante levar em conta a semântica dos locais, que pode ser obtida com a descrição disponível nas LBSNs, bem como o perfil dos indivíduos que frequentam o local, que poderia ser obtido, por exemplo, através das semelhanças entre usuários. A Figura \ref{figFoursquare} mostra um exemplo ilustrativo de recomendação de locais fornecida pelo Foursquare. Esse recomendador utiliza, além de outras informações, a opinião dos usuários que já visitaram previamente os estabelecimentos, bem como a popularidade do local e tendências que ocorrem na área estudada \cite{foursquareRec}.

Explorando também os dados do Foursquare, Mueller et al.~\cite{Mueller2017} propuseram uma metodologia para caracterizar as preferências de gênero por locais em diferentes regiões e em diferentes granularidades espaciais. Tradicionalmente, os dados utilizados para apoiar esses estudos são obtidos manualmente, muitas vezes por meio de pesquisas com voluntários. 
Os resultados desse estudo sugerem que a metodologia proposta pode ser uma ferramenta promissora para apoiar estudos sobre preferências de gênero por locais ao redor do mundo, sendo mais rápida e barata que os métodos tradicionais. 
Isso ilustra que os mesmos dados de LBSNs podem ser usados para diferentes propósitos.

\begin{figure}[!htbp]
\centering
\includegraphics[width=0.60\textwidth]{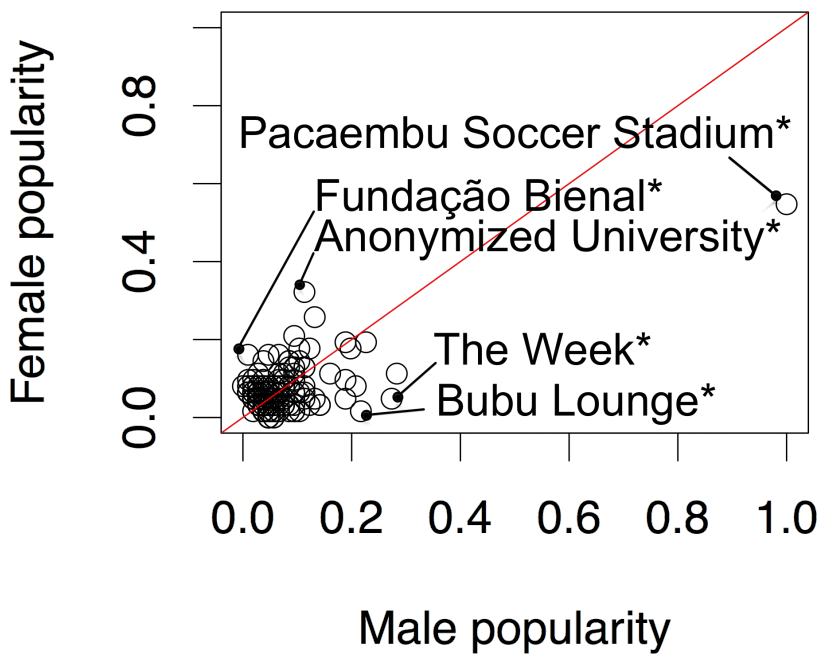}
\caption{Popularidade (normalizada) de locais individuais dentro de cada grupo de gênero em São Paulo, considerando todos os valores de todas as subcategorias. [Imagem de \cite{Mueller2017}].}
\label{figPrefGender}
\end{figure}

A Figura \ref{figPrefGender} exemplifica o resultado da metodologia proposta, mostrando que ela pode ajudar a identificar as preferências de gênero para locais com granularidades mais finas, por exemplo, considerando todos os locais individualmente para São Paulo. A metodologia proposta identificou 21 locais onde a diferença de popularidade entre gêneros é estatisticamente significativa. Um exemplo é uma universidade privada (que pediu explicitamente para ser anonimizada), que é mais popular entre usuários do sexo feminino. Isso pode ser explicado pela presença maior de mulheres nos cursos específicos localizados no campus (ou seja, saúde, artes, pedagogia e produção de mídia) no Brasil. Um porta-voz da universidade anônima confirmou, por e-mail, que a maioria pessoas matriculadas no campus são mulheres. Outro exemplo é o Museu de Arte da Fundação Bienal Ibirapuera, que também é significativamente mais popular entre mulheres. Este resultado foi confirmado por dados oficiais \cite{Mueller2017}.

Ao explorar as preferências de gênero por diferenças de locais, os proprietários de empresas poderiam obter informações valiosas sobre seus clientes. Além disso, as recomendações de locais podem se tornar mais conscientes culturalmente, já que homens e mulheres podem ter preferências diferentes em regiões com culturas distintas. Podemos citar ainda que estudos sociológicos sobre preferências de gênero por locais poderiam ser feitos em menos tempo, com tamanhos amostrais maiores e em regiões com granularidades espaciais arbitrárias \cite{Mueller2017}.

Para mais detalhes da utilidade de redes sociais baseadas em localização para área de computação urbana consulte \cite{Silva2018}. Nesse trabalho os autores discutem conceitos fundamentais da computação urbana utilizando dados do LBSN e apresentam um levantamento de estudos recentes de computação urbana que utilizam dados do LBSN.

\subsection{Aplicações e serviços em segurança pública}
\label{secSegPub} 

Segurança pública é exemplo de aplicação (ou serviço) desenvolvido no contexto da computação urbana. Sabe-se que alguns eventos observados em grandes centros urbanos, por exemplo, acidentes, congestionamentos, inundações e ataques terroristas, representam ameaças que deturpam a ordem das cidades, bem como apresentam riscos à segurança pública. Além disso, em tais centros, há uma ampla disponibilidade de diferentes tipos de dados que podem ser utilizados pela aplicação. Tanto para lidar corretamente com as ameaças mencionadas anteriormente, quanto para detectar uma ameaça ou até mesmo prevê-las~\cite{riveiro2017anomaly,wang2017road,akabane2018imob}.

\begin{figure}[!htbp]
\centering
\includegraphics[width=.95\textwidth]{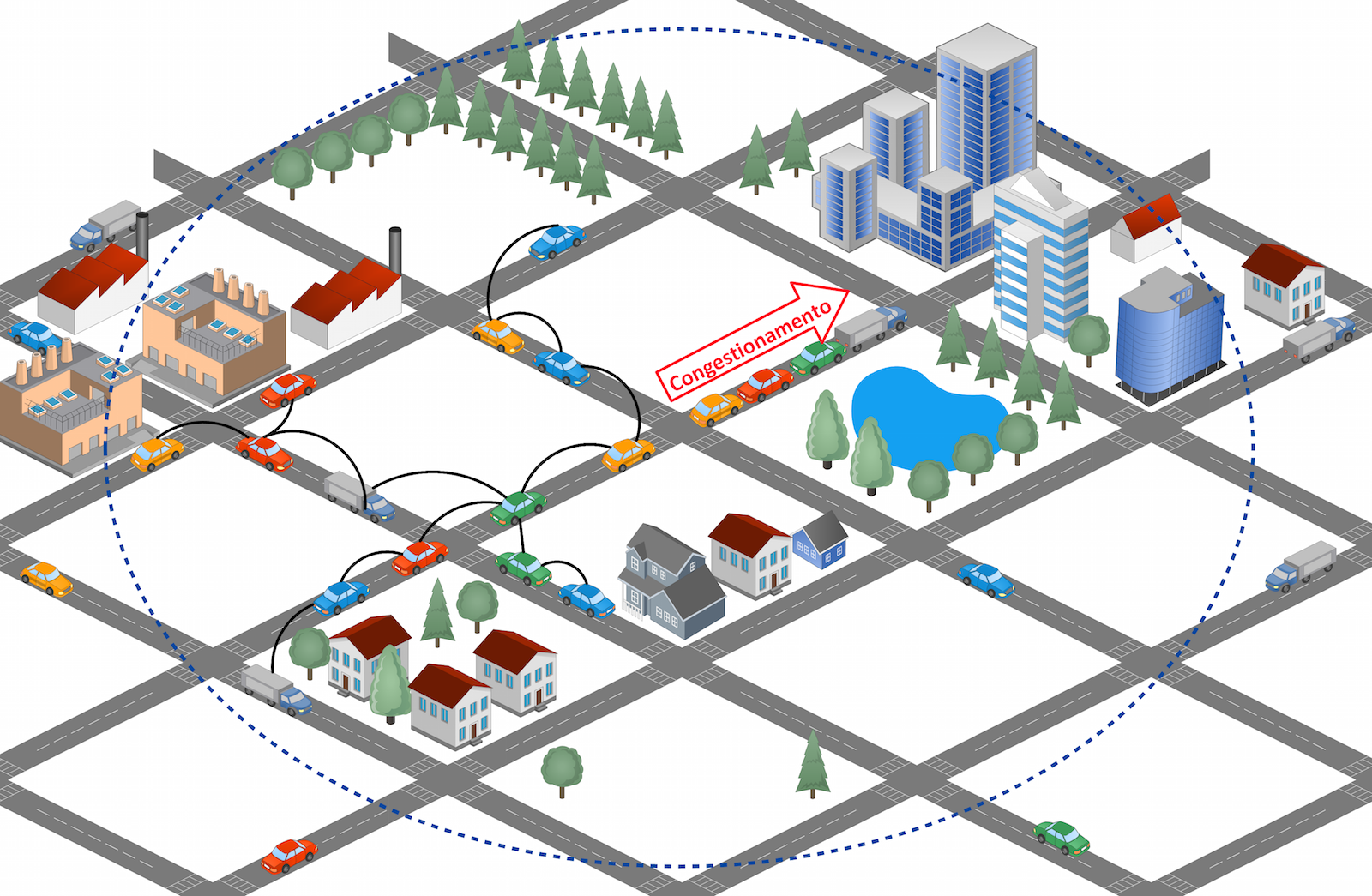}
\caption{Nesse exemplo ilustrativo o veículo mais próximo do congestionamento escolhe primeiro uma rota alternativa e a repassa para a sua vizinhança. O segundo mais próximo utiliza tal rota para calcular a sua e assim por diante. \cite{akabane2018imob} - modificado.}
\label{fig:congestionamento}
\end{figure}

Além da segurança pública, outra aplicação de grande importância dentro da computação urbana é a detecção de anomalias. Uma das anomalias comumente observada em áreas urbanas é de tráfego, particularmente congestionamento. Tal anomalia pode ser causada por diversos fatores, por exemplo, acidentes, protestos, construção e desastres naturais. Assim a detecção ou prevenção de tais anomalias podem ajudar na gestão de tráfego. Algumas soluções podem ser encontradas em trabalhos relacionados que lidam com a detecção de tal anomalia~\cite{de2017traffic,riveiro2017anomaly,wang2017road,akabane2018imob}. Dentre as soluções mencionadas anteriormente, apenas os trabalhos de \cite{de2017traffic} e \cite{akabane2018imob} propõem eliminar (ou amenizar) a anomalia de forma reativa. Em outras palavras, uma vez identificada a anomalia de tráfego veicular, cada solução aplica uma estratégia de re-roteamento dos veículos que estão deslocando em direção ao congestionamento. A principal diferença entre as duas soluções é que no trabalho de \cite{akabane2018imob} a etapa do cálculo da rota alternativa é realizada de forma colaborativa. Isto é, tal cálculo é realizado com base nas rotas escolhidas pela vizinhança, além do mais os veículos mais próximos do local do congestionamento têm a prioridade na escolha de uma rota alternativa, veja a Figura \ref{fig:congestionamento}.

Outra aplicação importante dentro de segurança pública é o de detecção de desastre naturais e evacuação. Sabe-se que o grande terremoto que atingiu o leste do Japão\footnote{https://www.livescience.com/39110-japan-2011-earthquake-tsunami-facts.html} e o acidente nuclear de Fukushima\footnote{https://www.britannica.com/event/Fukushima-accident} (também no Japão) causaram grandes movimentos populacionais e evacuações na região afetada. Logo, compreender e prever tais movimentos e eventos naturais são fundamentais para o planejamento de ajuda humanitária, gestão de desastres e reconstrução social de longo prazo~\cite{zheng2014urban}.

Seguindo essa ideia, no trabalho de \cite{song2013modeling} foi construído um grande banco de dados contendo registros de mobilidade humana coletado do GPS dos \textit{smartphones}. Tal banco possui as mobilidades de mais 1,6 milhões de pessoas que foram realizadas entre os dias de 1º de agosto de 2010 a 31 de julho de 2011. Após o tratamento dos dados foi aplicado um modelo probabilístico que encontrou padrões de comportamentos de evacuação de curto e longo-prazo da população após o desastre. Por meio desse modelo foi possível inferir a mobilidade de evacuação da população de outras cidades japonesas. Em outro trabalho, os autores~\cite{yabe2016framework} propuseram um arcabouço para estimar com eficiência os pontos críticos de evacuação após grandes desastres naturais em região urbana usando dados de localização coletados dos~\textit{smartphones}.

\section{Desafios e Oportunidades} 
\label{sec:desafios}

Nesta seção serão apresentados os principais desafios enfrentados pela evolução da computação urbana e exploração de novas aplicações para este cenário, destacando oportunidades de pesquisa que podem surgir a partir das ideias apresentadas. 

\subsection{Aquisição e Análise dos Dados} 

Sob o ponto de vista da aquisição dos dados em ambientes urbanos, diversos desafios podem ser enumerados. Como primeiro deles, a grande variedade dos diferentes modelos de dispositivos. Muitos são os dispositivos e sensores instalados em uma cidade, provenientes de serviços governamentais ou mesmo privados, que geram dados para diferentes propósitos. Além disso, diversos outros aplicativos de \textit{smartphones} são usados por usuários para acessar serviços e coletar dados diariamente. Assim, a integração desses dados provenientes de diferentes plataformas com o objetivo de prover novos serviços para os usuários torna-se um grande desafio.

Outro aspecto importante na aquisição dos dados está nas diferenças estruturais presentes nos dados coletados. Questões relativas a granularidade, escalas, precisão dos dados podem dificultar todo o processo de análise e interpretação dos mesmos. Além disso, todo o processo de coleta pode apresentar anomalias e até mesmo algumas falhas, o que pode provocar lacunas no monitoramento realizado. Em situações como essas, o emprego de técnicas estatísticas e modelos matemáticos são fortemente necessários para remover as diferenças entre os dados de forma a obter uma base de dados mais coesa e confiável~\cite{celes2017}.  

O aspecto localidade é outro problema na análise de dados em áreas urbanas. Devido às características de conectividade, concentração de pessoais e de estabelecimentos comerciais, lugares mais populares tendem a ser mais monitorados. Por exemplo, o aplicativo Instagram\footnote{http://www.instagram.com} torna-se uma grande fonte de dados acerca do padrão de visitação turística de uma cidade. Entretanto, quando se necessita analisar o padrão de toda a cidade, áreas mais periféricas, afastadas do centro normalmente possuem poucos registros de dados coletados. Nesta circunstância, a correlação de dados de diferentes fontes se torna muito necessária no estudo destas regiões. 

Neste mesmo contexto, o armazenamento dos dados também se torna um grande desafio. Alguns dispositivos eletrônicos, devido a suas peculiaridades, apresentam pouca memória, o que demanda o uso de técnicas de compressão e organização dos dados de forma a garantir um melhor gerenciamento de todas as variáveis monitoradas. Assim, algoritmos específicos para trabalhar com os dados coletados e suas diferenças, são essenciais nestes cenários restritivos. Em outra direção, a fusão de dados, a filtragem e agregação podem ser uma boa estratégia para reduzir a quantidade de dados armazenados e facilitar a análise e indexação dos mesmos~\cite{zheng2015}. 

Levando em consideração os dados pessoais, a privacidade passar a ser uma grande barreira para sua coleta e análise. Em muitos cenários, a identificação do dado é um fator restritivo para coleta, impedindo que o usuário permita o acesso e o compartilhamento desses dados. Dessa forma, os serviços que fazem uso de dados pessoais devem criar mecanismos para garantir segurança aos seus usuários, incentivando o seu uso. Em outra abordagem, na literatura pode-se encontrar muitos trabalhos que apresentam discussões sobre privacidade e segurança de dados em ambientes urbanos, que apresentam técnicas de anonimização dos dados evitando essa identificação~\cite{ji2017}. Em contrapartida, novos projetos de leis são propostos pelo governo no intuito de suprir essa lacuna, dando maior proteção aos usuários.  

Tendo em vista a análise dos dados urbanos, as diferentes características presentes nas cidades podem dificultar o processo de modelagem e análise dos dados. Cidades apresentam características peculiares, dentre elas pode-se citar: relevo, clima, ocupação do solo, padrões de mobilidade, etc. E a modelagem destes dados se torna uma tarefa não trivial. Em geral, modelos são criados a partir de padrões, e a variação destes padrões tende a inviabilizar a aplicação de um mesmo modelo para representar comportamentos de diferentes ambientes urbanos~\cite{silvaUserUrbSensing}. Assim, novas oportunidades de estudos e definições de modelos que se melhor adéquem a essa diversidade de características se torna essencial no cenário da computação urbana. 

\subsection{Desenvolvimento de aplicações} 

Como pode ser observado na Seção~\ref{sec:aplicacoes}, existem diversas aplicações relacionadas ao âmbito de Computação Urbana. Nada obstante, vários desafios são adicionados devido as exigências de qualidade impostas por cada classe de aplicação, em vista disso será listado os principais desafios relacionado aos tipos abordados na Seção.

Dentro do contexto de Sistemas de ITSs, existem os desafios relacionados a maneira eficiente em que estes sistemas irão atuar, visando garantir a segurança dos usuários. A arquitetura dos ITSs possui diversos componentes que garantem a comunicação entre os dispositivos, podendo adotar várias tecnologias, tais como DSRC, Wi-Fi, LTE, Visible Light Communication, satélite \cite{zheng2015heterogeneous}. Portanto, pode-se considerar um desafio projetar as comunicações apropriadas neste cenário heterogêneo de tecnologias.

Dados coletados de sistemas de transporte costumava ser dados não pessoais, como dados de fluxo de tráfego e localização de veículos, porém dados pessoais podem ser necessários durante o processamento de algumas aplicações. Nesse contexto, a privacidade é uma das tarefas mais desafiadoras dentro das aplicações de ITS \cite{zhu2018big}. Os departamentos de transporte devem regulamentar rigorosamente a definição de dados pessoais utilizados nas aplicações, além de fortalecer o gerenciamento da segurança de dados e aplicar algoritmos para melhorar o nível de segurança dos dados coletados.

Os ITS ainda precisam lidar com a dinâmica das cidades, vários dados estão disponíveis para analisar o comportamento das cidades. Dessa forma, a redução de congestionamentos, poluição sonora e ambiental, além de proporcionar uma locomoção mais segura e eficiente são os principais objetivos desse tipo de sistema \cite{cunha2017sistemas}. Os dados de mobilidade ajudam a assimilar a rotina das pessoas e assim fornecer aplicações que otimizam os meios de transporte.

Outra preocupação recorrente, está relacionada a coleta dos dados, uma vez que os veículos são equipados com uma grande quantidade de sensores. Estes sensores coletam dados que são utilizados para a tomada de decisão dos sistemas de controle de direção. Assim, o principal desafio é extrair as informações desses sensores e obter um conhecimento a partir desses dados, além de relacionar as informações obtidas com dados externos aos veículos. 

A predição de tráfego é um dos desafios que influencia no comportamento das aplicações, os motoristas podem solicitar informações sobre a condição das estradas, com o objetivo de evitar possíveis congestionamentos e acidentes. A predição do trafego pode ser realizada com uma cooperação entre veículos, criando uma \textit{Vehicular Fog Computing} (VFC), responsável por sensoriamento, processamento e compartilhamento de informações \cite{ning2019vehicular}.

Além de que existem aplicações gerando uma grande quantidade de dados, encontra-se a necessidade de realizar o \textit{offloading} de maneira eficiente. Uma VFC pode prover esse tipo de serviço buscando reduzir a latência e o custo de disseminação de uma determinada aplicação \cite{zhang2017mobile}. Um mecanismo de \textit{offloading} preditivo pode ser projetado, ao estimar o consumo de tempo, para decidir se o envio direto ou o encaminhamento de retransmissão preditiva é adequado para comunicações veiculares. Já no caso de aplicações sociais baseadas em localização a quantidade e a confiabilidade dos dados inseridos pelos usuários são os principais pontos que devem ser examinados. Os dados disponíveis para as aplicações podem representar apenas uma parte dos hábitos dos usuários e a disponibilidade desses dados pode ser afetada por fatores externos, como condições climáticas. 

Além disso, não se pode apenas assumir que os dados inseridos pelos usuários estão totalmente corretos, algumas informações podem ser geradas deliberadamente pelos usuários, o que pode introduzir uma inconsistência no que é aferido. Um exemplo disso pode acontecer quando um determinado usuário realiza um \textit{check-in} falso em um restaurante por estar na moda, o que influenciaria no sistema de recomendação ou popularidade de um determinado local. É importante destacar a etapa de coleta desses dados, uma das dificuldades está no desenvolvimento de ferramentas para coletas, levando em consideração a eficiência, uma vez que se trata de uma grande quantidade de dados provenientes de fontes distintas. Esse mecanismo deve lidar ainda com questões de restrição de privacidade envolvendo os usuários. Além de garantir que os dados coletados sejam verdadeiros\cite{santos2017towards}.

Outra dificuldade encontrada nesta área está relacionada as múltiplas fontes de dados urbanos, em aplicações desse tipo é necessário extrair informações de diversas fontes, com dados distintos (Por exemplo, vídeos, imagens e textos). Os algoritmos devem lidar com uma grande quantidade de dados gerados por vários tipos de LBSN, dessa forma, se faz necessário saber como integrar essas fontes heterogêneas de dados, como agregar ou até mesmo desenvolver algoritmos para filtrar tais dados. 

A localização desempenha um papel importante nas aplicações citadas, sendo um dos seus principais problemas a ser resolvidos: Como prover dados de posicionamento a qualquer momento e lugar de uma forma mais acurada e confiável? No cenário da computação urbana, a natureza do deslocamento em alta velocidade dos veículos causa mudanças rápidas e constantes na topologia das redes veiculares, o que provoca a disseminação de informações desatualizadas, esse é outro ponto que deve ser levado em consideração quando são propostas aplicações neste sentido.

Aplicações críticas, como as associadas a veículos autônomos, podem exigir dados de localização com baixas taxas de erro. Erros de localização em GPS estão em torno de 10 a 30 metros\cite{nascimento2018integrated}, o que cria a necessidade de aplicar técnicas para reduzir esses erros, como \textit{Map Matching} e \textit{Dead Reckoning}. Entretanto, ainda é um desafio pertinente prover serviços de localização com uma alta acurácia e em todos os cenários, como tuneis por exemplo\cite{golestan2015localization}.

Por fim, nos casos das aplicações relacionadas a segurança urbana, existe a necessidade de levar em consideração os dados disponíveis para propor as rotas mais seguras para os usuários. A maioria dos sistemas propostos foca apenas no menor caminho possível, podendo levar os veículos a lugares perigosos. Nesse contexto é necessário unir os dados de diferentes fontes para achar a melhor rota que combine segurança e tempo. 

Outro desafio trata como fazer o cálculo de rotas seguras e eficientes sem adicionar um tempo maior de processamento e evitar congestionamentos. Esse fator depende do número de entidades que serão roteadas (carros, motos, ônibus, etc), o que pode aumentar o tempo necessário para processar todas as informações coletadas e sugerir as melhores rotas\cite{de2017traffic}.

Existem outros desafios relacionados a aplicações específicas e podem ser citados como o processamento em tempo real de grandes quantidades de dados, interação com usuários ou um grupo de usuários, a interação com o ambiente em que estas aplicações estão inseridas, além de uma falta de \textit{testbeds}, o que torna difícil compreender o real comportamento e desafios das aplicações \cite{santana2018software}. 

\section{Um Estudo de Caso: Computação Urbana na Prática}
\label{sec:estudodecaso}

A fim de aumentar a familiaridade da audiência com o arcabouço de computação urbana, apresentamos um estudo de caso que explora a utilização de dados urbanos disponíveis em plataformas online publicamente para identificar fluxos de mobilidade nas cidades, com o intuito de oferecer opções de rotas multimodais para os cidadãos. Para isso, utilizamos como cenário a cidade de São Paulo, SP, motivado por diversos fatores, tais como problemas recorrentes de congestionamento que afeta diariamente a vida dos pessoas que moram ou visitam São Paulo e a necessidade das pessoas em percorrer grandes distâncias para deslocarem-se entre suas casas e trabalhos (ou escolas, lugares de lazer, etc), além de haver uma grande quantidade de dados disponíveis que são cruciais para aplicação.

Esta seção está organizada como se segue. A Subseção \ref{subsec:estudodecaso:fluxos} apresenta o arcabouço utilizado para identificação de fluxos de mobilidade em São Paulo. Baseado nos fluxos identificados, a Subseção \ref{subsec:estudodecaso:routing} descreve o passo chave da aplicação, o algoritmo de sugestão de rotas multimodais, que além das rotas multimodais também calcula rotas ``tradicionais'', i.e., que consideram apenas transporte público ou apenas veículo privado como meio de transporte, para possibilitar a realização de uma análise comparativa entre as rotas sugeridas. Como o custo financeiro para percorrer uma rota com um veículo privado pode variar de acordo com o tipo de veículo, simplificamos ao considerar que é utilizado o serviço de transporte da empresa Uber \textit{Technologies Inc}., assim, os resultados apresentados utilizam a palavra ``Uber'' para indicar que o meio de transporte considerado é o veículo privado. Também é apresentado um algoritmo para geração de visualização das rotas sugeridas em mapas 2D interativos. A Subseção \ref{subsec:estudodecaso:results} demonstra a vantagem em se utilizar as rotas multimodais em comparação as rotas tradicionais ao conseguir um bom equilíbrio em relação as métricas de custo financeiro (preço), tempo de viagem estimado (duração) e distância percorrida a pé. Todos os materiais, algoritmos, ferramentas e tutoriais para instalação e execução da aplicação pode ser acessado no Github\footnote{https://github.com/diegopso/hybrid-urban-routing-tutorial-sbrc}.

\subsection{Identificação de Fluxos Urbanos} \label{subsec:estudodecaso:fluxos}
Como já supracitado, há várias maneiras para extração da mobilidade das pessoas nas cidades (i.e., os fluxos urbanos) a partir de dados urbanos. Entre elas, temos o SMAFramework \cite{RODRIGUES2018111}, um arcabouço para análise de dados de mobilidade urbana que disponibiliza um algoritmo que permite o agrupamento de viagens urbanas em fluxos de mobilidade e classifica esses fluxos em tendências e secundários. O SMAFramework utiliza o algoritmo HDBSCAN para identificar zonas funcionais nos inícios e finais de viagens da base de dados, utilizadas para computação e classificação de fluxos entre as zonas funcionais. Aqui, o SMAFramework é utilizado como uma ``caixa preta'' para identificação dos principais fluxos, onde é detalhado o processo para aquisição e processamento da instância de entrada para o algoritmo e a utilização da saída gerada (os fluxos urbanos), mas o funcionamento do SMAFramework é explicado em alto nível. É recomendado que o leitor explore não apenas o SMAFramework, mas também outras abordagens que efetuem a identificação de fluxos urbanos (ou desenvolva sua própria solução), os quais podem ser facilmente adequados para serem utilizados em conjunto com essa aplicação.

\subsubsection{Aquisição de dados} \label{subsubsec:estudodecaso:data-acquisition}
O primeiro passo é coletar uma grande quantidade de dados urbanos que possibilitem a identificação da mobilidade urbana. Como visto anteriormente, as redes sociais baseada em localização (LBSNs) são uma rica fonte de dados geográficos compartilhados pelos usuários em tempo real. Por isso, é utilizado o Twitter como fonte de dados, que é uma grande LBSN e oferece aos usuários o serviço de \textit{microblogging} \textit{online}, onde eles podem compartilhar mensagens curtas (máximo de 280 caracteres) conhecidas como \textit{tweets}. Por meio da API do Twitter é possível coletar \textit{tweets} de áreas de interesse, onde parte desse conjunto de \textit{tweets} é geo-localizado. Como a maioria das APIs, a API do twitter também impõem aos seus utilizadores uma série de restrições\footnote{https://developer.twitter.com/en/docs/basics/rate-limiting.html} que devem ser consideradas durante o processo de aquisição de \textit{tweets}.

Um objeto \textit{tweet} \footnote{https://developer.twitter.com/en/docs/tweets/data-dictionary/overview/tweet-object.html} coletado por meio da API (renderizado em JSON), é composto por uma longa lista de atributos como  o $id$ do \textit{tweet}, o conteúdo da mensagem ($text$), informações sobre o usuário ($user$), o tempo em que o \textit{tweet} foi criado em mili-segundos ($timestamp\_ms$, onde o fuso horário é UTC), as coordenadas do local se o \textit{tweet} é geolocalizado ($coords$), entre outros. Em geral, o tamanho de um \textit{tweet} renderizado é cerca de 5KB, mas como são coletados centenas de milhares de \textit{tweets}, a quantidade de espaço para armazenamento pode facilmente extrapolar as dezenas e até centenas de GB. Por isso, o conjunto de dados de \textit{tweets} geo-localizados disponibilizado para o minicurso (chamado de \url{geotagged\_tweets\_sp.csv}), contém apenas os atributos requeridos pelo SMAFramework, a saber: um \textit{id} aleatório para o usuário ($uid$), as coordenadas (latitude -- $lat$ -- e longitude -- $lon$) e o tempo de criação do \textit{tweet} ($timestamp\_ms$), o que reduz drasticamente a quantidade de espaço requerida (de $\pm$3,3GB para $\pm$44MB).

\subsubsection{Relacionamento e Filtro de Dados} \label{subsubsec:estudodecaso:record-linkage}

De posse do \textit{dataset} \url{geotagged\_tweets\_sp.csv}, o próximo passo consiste em relacionar os \textit{tweets} existentes em fluxos urbanos. Para isso, o primeiro passo é agrupar os \textit{tweets} de acordo com o atributo $uid$, onde cada grupo é composto por \textit{tweets} compartilhados por um mesmo usuário. Então, para cada grupo são criados \textit{links} entre os \textit{tweets}, indicando uma sequência temporal entre os \textit{tweets}. Por exemplo, suponha que o usuário cujo $uid=1$ tenha compartilhado 10 \textit{tweets} ao longo de um dia, denotado por $tweet_{uid=1}^0, tweet_{uid=1}^1, tweet_{uid=1}^2, ..., tweet_{uid=1}^9$. Assim, um \textit{link} entre dois \textit{tweets}, por exemplo, $tweet_{uid=1}^i \rightarrow tweet_{uid=1}^j$, onde $0 \leq i \neq j < 10$, indica que o \textit{tweet} $tweet_{uid=1}^j$ foi compartilhado subsequente ao \textit{tweet} $tweet_{uid=1}^i$, sendo os \textit{tweets} $tweet_{uid=1}^i$ e $tweet_{uid=1}^j$ contíguos nessa sequência temporal.

Note que ao criar tais \textit{links}, algumas dessas conexões podem não ser interessantes para nossa aplicação. Por exemplo, conexões entre dois \textit{tweets} que foram compartilhados em dias diferentes potencialmente não indica um fluxo válido, uma vez que o usuário pode ter se deslocado para diferentes áreas na janela temporal entre os dois \textit{tweets}, não implicando em fluxo direto entre as localizações em que os \textit{tweets} foram compartilhados. Para evitar esse tipo de ruído e alguns outros, foram definidos e utilizados filtros, como apresenta a Tabela \ref{tab:estudodecaso:filtros}. Essa tabela também mostra para cada passo, a descrição e a quantidade de dados restantes no conjunto de dados, sendo que para o passo Inicial (linha 1), a quantidade é igual o número total de \textit{tweets} e, para as demais linhas, a quantidade é referente ao total de conexões.

\begin{table}[!htbp]
\centering
\footnotesize{
\begin{tabular}{cccc}
\cline{2-4}
 & \textbf{Passo} & \textbf{Descrição} & \textbf{Quantidade} \\ \cline{2-4} 
\multirow{2}{*}{} & Inicial & \begin{tabular}[c]{@{}c@{}}Todos os \textit{tweets} analisados,\\ ainda não agrupados\end{tabular} & 690139 \\
 & Conexão de \textit{tweets} & \begin{tabular}[c]{@{}c@{}}\textit{Tweets} conectados em\\ sequências temporais\end{tabular} & 588760 \\ \cline{2-4} 
\multirow{4}{*}{\rotatebox[origin=c]{90}{\textbf{Filtros}}} & Seleção de fluxos diários & \begin{tabular}[c]{@{}c@{}}Filtro de conexões com datas\\ inicial e final distintas\end{tabular} & 94411 \\
& Restrição de distância mínima & \begin{tabular}[c]{@{}c@{}}Filtro de conexões com distância\\ percorrida menor que 100m\end{tabular} & 40612 \\
& Restrição de duração mínima & \begin{tabular}[c]{@{}c@{}}Filtro de conexões com duração\\ menor que 1 segundo\end{tabular} & 40567 \\
& Restrição de velocidade & \begin{tabular}[c]{@{}c@{}}Filtro de conexões com velocidade menor\\ que 2 Km/h e maior que 100 Km/h\end{tabular} & 15675 \\ \cline{2-4} 
\end{tabular}
}
\centering
\caption {Passos para seleção de fluxos válidos.}
\label{tab:estudodecaso:filtros}
\end{table}

\begin{figure}[!htbp]
    \centering
        \includegraphics[width=.75\textwidth]{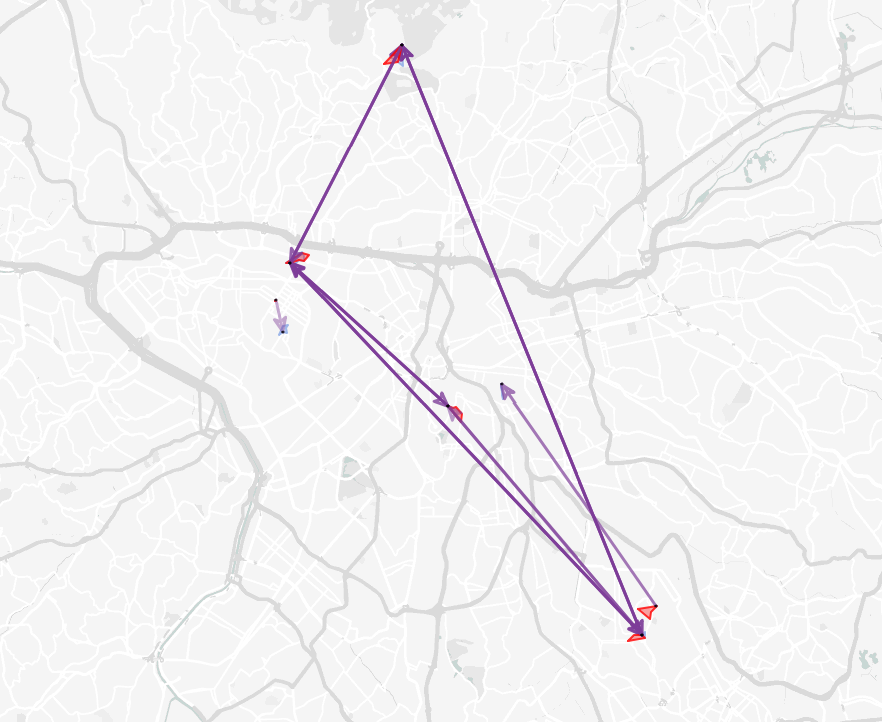}
        \caption{Principais fluxos identificados.}
    \label{fig:estudodecaso:fluxos}
\end{figure}

Feito isso, a instância de entrada para o algoritmo do SMAFramework está pronta. A partir de todos os fluxos identificados, o SMAFramework determina os principais fluxos de mobilidade, como mostra a Figura \ref{fig:estudodecaso:fluxos}. Ao todo, foram identificados 12 fluxos de mobilidade urbana em São Paulo, SP, a partir dos \textit{tweets} analisados. Como podemos observar, a maioria desses fluxos estão concentrados em quatro regiões, que são os bairros: Jardim Itatinga, Olímpico em São Caetano do Sul, Jardim das Perdizes e Aclimação.
Por isso, somente os fluxos entre esses bairros serão considerados para os próximos passos da aplicação (um total de 7 fluxos). Alternativamente, todos os fluxos identificados, além dos 12 resultantes do SMAFramework, poderiam ser avaliados individualmente, contudo, tal análise demandaria muitos recursos computacionais e quantidade de tempo.

\subsection{Rotas Multimodais} \label{subsec:estudodecaso:routing}
Para percorrer um dado fluxo, podemos considerar três opções de meio de transporte diferentes, são eles: a pé, transporte público (i.e., metrô e ônibus) e veículo privado (nesse caso, Uber). Assim, para cada fluxo de viagem, são calculadas rotas e gerados mapas considerando
os seguintes tipos de mobilidade: prioritariamente com transporte público (rotas de cor vermelha), onde alguns trechos da rota possivelmente são realizados a pé (rotas em verde) devido a transição entre as estações; utilizando o serviço de Uber; e as rotas multimodais, onde as três opções de meio de transporte podem ser combinadas de diferentes maneiras. Aqui, apresentamos duas combinações possíveis de rotas multimodais, uma sendo prioritariamente com o transporte público, denotada por ``Híbrida 1'', e a outra sendo prioritariamente com o Uber, denotada por ``Híbrida 2''. No entanto, é importante ressaltar que os algoritmos desenvolvidos calculam e geram diversas opções de rotas multimodais. Além disso, como a maioria dos fluxos identificados são demasiadamente longos para serem percorridos totalmente a pé, esse meio de transporte é apenas considerado em conjunto com outros meios de transporte.

\subsubsection{Cálculo de Rotas} \label{subsubsec:estudodecaso:computing_routes}
Dado os pontos de origem e destino de algum fluxo, o Algoritmo \ref{alg:compute-hybrid-routes} descreve os principais passos para calcular diversas opções de rotas multimodais. Inicialmente, o algoritmo inicializa as seguintes variáveis (linhas 1--4): $driveng\_way$ que armazena uma rota entre a origem e o destino, considerando o meio de transporte como sendo via veículo privado (carro), e é utilizada para determinar as condições de tráfego no trajeto. Note que várias opções de rota podem ser utilizadas como ponto de partida para avaliar outros trajetos com mesma origem e destino. Aqui, essa rota foi determinada por meio da TomTom Routing API\footnote{https://developer.tomtom.com}; as listas $transit\_start\_candidates$ e $transit\_end\_candidates$, que vão armazenar os pontos para transição de meios de transporte; A lista $options$ que irá armazenar todas as opções de rotas multimodais. É importante ressaltar que o início e o fim da rota também foram adicionados como candidatados à transição (linhas 5--6), isso é feito para obter-se também as rotas tradicionais como opções.

Feito isso, o algoritmo percorre todos as instruções na rota original, i.e., as orientações fornecidas aos motoristas para chegar ao seu destino, com o intuito de identificar os trechos congestionados na rota e adicionar seus inícios e finais às listas de pontos de transição (laço das linhas 7--12). Uma vez identificados os pontos de transição, o algoritmo realiza as combinações candidatas de início e fim -- $ts,\; te$ -- para compor diferentes rotas multimodais com o apoio das APIs de cálculo de rotas do Google Directions, TomTom Routing e Uber. A API do Google Directions foi usada para avaliar as possíveis rotas, ou trechos, que utilizam transporte público; essa API considera todas as opções de transporte público existente, como metrô, ônibus, tram, etc. A API TomTom Routing é utilizada para calcular a rota referência para selecionar os pontos de transição; ela foi escolhida pois junto a rota traz informações sobre o congestionamento das vias. Por fim, a API do Uber foi utilizada para estimar o valor e o tempo de espera das rotas que utilizam esse meio de transporte. Todas as opções são concatenadas a lista de opções da rota a fim de serem retornadas como saída do algoritmo (laço das linhas 13--18).

\begin{algorithm}[!ht]
 
 \caption{Encontrar opções de rotas multimodais.}
 \label{alg:compute-hybrid-routes}
 
 \SetKwInOut{Input}{Entrada}\SetKwInOut{Output}{Saída}
 \Input{Origem e destino de um fluxo $(origin, destination)$.}
 \Output{Lista com as opções de rotas multimodais.}
 
 \BlankLine
 
 $\textit{drive\_way} \gets get\_driving\_way (origin,\, destination)$ \\
 $\textit{transit\_start\_candidates } \gets new\, List()$ \\
 $\textit{transit\_end\_candidates} \gets new\, List()$ \\
 $\textit{options} \gets new\, List()$ \\
 
 \BlankLine
 
 $append(transit\_start\_candidates,\, origin)$ \label{line:append-origin} \\ 
 $append(transit\_end\_candidates,\, destination)$ \label{line:append-destination} \\
 
 \BlankLine
 
 \ForEach  {$(index, step)~\textbf{in}~drive\_way.steps$} { \label{line:foreach-driveways}
  \If {$is\_congested(step)$} {
     $append(transit\_start\_candidates,\, step.origin)$ \\
     $append(transit\_end\_candidates,\, step.destination)$ \\ 
  } \label{line:get-traffic-data-end} 
 } 
 
 \BlankLine
 
 \ForEach {$ts~\textbf{in}~transit\_start\_candidates$} { \label{line:hybrid-mixed-start} 
  \ForEach {$te~\textbf{in}~transit\_end\_candidates$} {
     $options \gets get\_options(origin,\, ts,\, te,\, destination)$ \\
     $concat(options, opts)$ \label{line:concat-3} 
  } 
 } \label{line:hybrid-mixed-end} 
 
\end{algorithm}

\subsubsection{Visualização de Rotas} \label{subsubsec:estudodecaso:visualization}
Com as opções de rotas calculadas para todos os fluxos estudados, falta gerar os mapas com as rotas sugeridas. O algoritmo \ref{alg:visualization} mostra os principais passos para geração de mapas 2D interativos. Inicialmente, o algoritmo inicializa o mapa (linha 1), que será gerado como saída com a rota sugerida (informada como entrada do algoritmo). Cada trecho da rota (chamado de \textit{step}) calculado pelo algoritmo \ref{line:hybrid-mixed-end}, possuí uma série de informações, tais como a origem, o destino, o tempo de viagem e, os mais importantes para geração da visualização, a \textit{overview polyline} e o modo de transporte. De acordo com o Google Maps\footnote{https://developers.google.com/maps/documentation/directions/intro}, a \textit{overview polyline} é uma representação codificada de um caminho aproximado da rota sugerida e, ao ser decodificada\footnote{https://developers.google.com/maps/documentation/utilities/polylinealgorithm}, é possível obter uma sequência de pontos (coordenadas) que formam a rota referente ao \textit{step}. Assim, o algoritmo primeiro decodifica a \textit{overview polyline} e armazena em uma lista de coordenadas (linha 3). Então, esta lista de coordenadas é plotada no mapa, onde a cor da linha que representa este caminho é definida de acordo com o meio de transporte e a largura da linha pode ser definida por um valor inteiro (linha 4). Ao final, todos os trechos da rota são plotados no mapa, compondo a rota total com a coloração indicando o meio de transporte em cada trecho. Como resultado, são apresentados a seguir os mapas com as quatro opções de rotas para cada fluxo.

\begin{algorithm}[!ht]
\SetKwInOut{Input}{entrada}\SetKwInOut{Output}{saída}
\Input{$route$.}
\Output{Mapa com a rota sugerida.}
\BlankLine
\tcp{Inicializa o mapa, de acordo com o cenário (neste caso, São Paulo).}
$map = setMap(central\_coord=(-23.551615,-46.633611), zoom=12)$; \\
\ForEach{$step \in route$ }
{
     \tcp{Decodifica a $overview\_polyline$}
     $path \leftarrow decode(step[``overview\_polyline"])$; \\
     \tcp{Adiciona o caminho ao mapa, onde a cor é definida pelo meio de transporte.}
     $map.plot(path, color=step[``travel\_mode"], edge\_width=3)$; \\
}
$map.draw(output.html)$
\caption{Visualização de rotas.}\label{alg:visualization}
\end{algorithm}

\begin{figure}[!htp]
    \centering
    \subfigure[Transporte público.]{
        \includegraphics[scale=0.23]{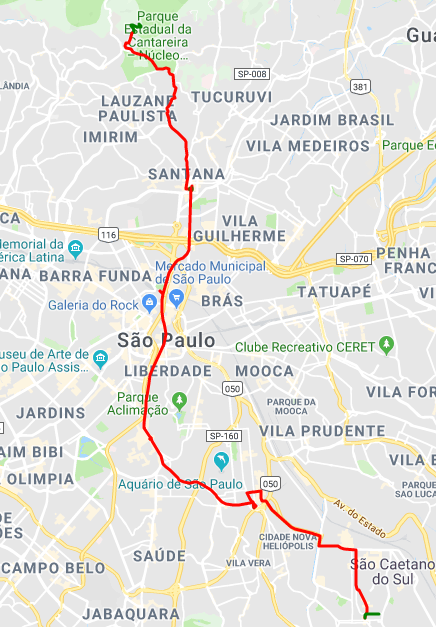}} 
    \subfigure[Uber.]{
        \includegraphics[scale=0.23]{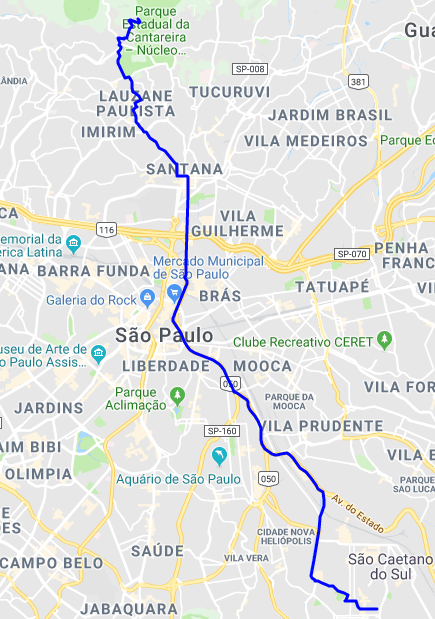}}
    \subfigure[Híbrida 1.]{
        \includegraphics[scale=0.23]{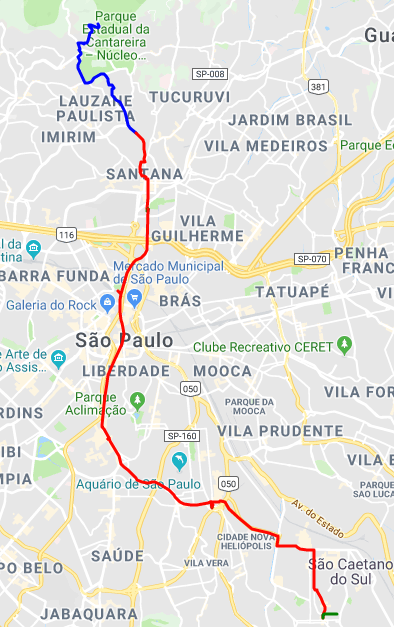}} 
    \subfigure[Híbrida 2.]{
        \includegraphics[scale=0.23]{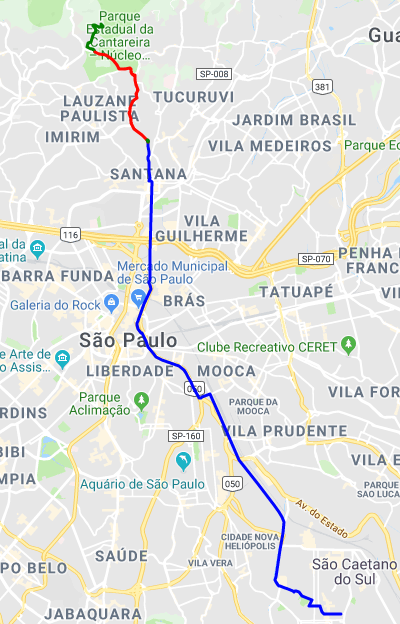}} 
    \caption{\textbf{Fluxo 1 --} Do bairro Jardim Itatinga para o bairro Olímpico, São Caetano do Sul.}
    \label{fig:map_fluxo_1}
\end{figure}

\begin{figure}[!htp]
    \centering
    \subfigure[Transporte público.]{
        \includegraphics[scale=0.195]{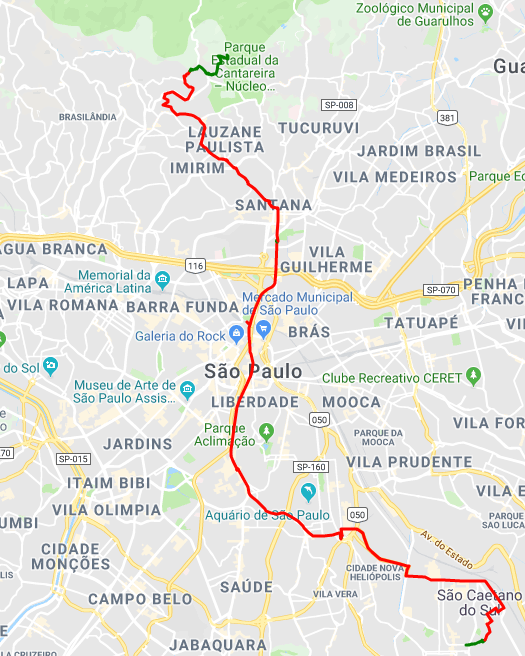}} 
    \subfigure[Uber.]{
        \includegraphics[scale=0.198]{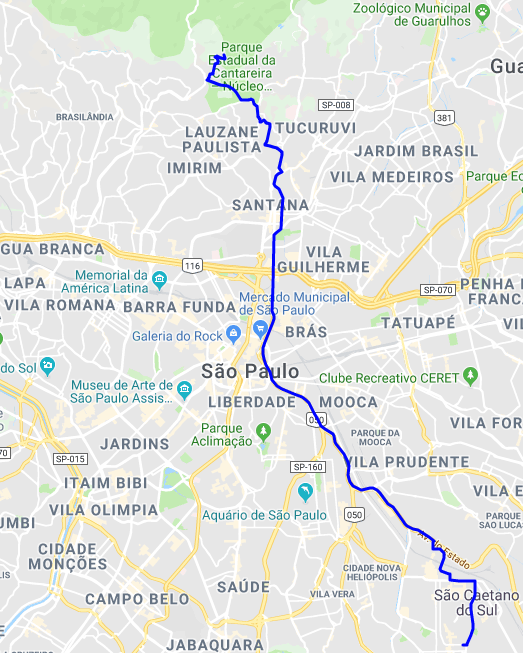}}
    \subfigure[Híbrida 1.]{
        \includegraphics[scale=0.183]{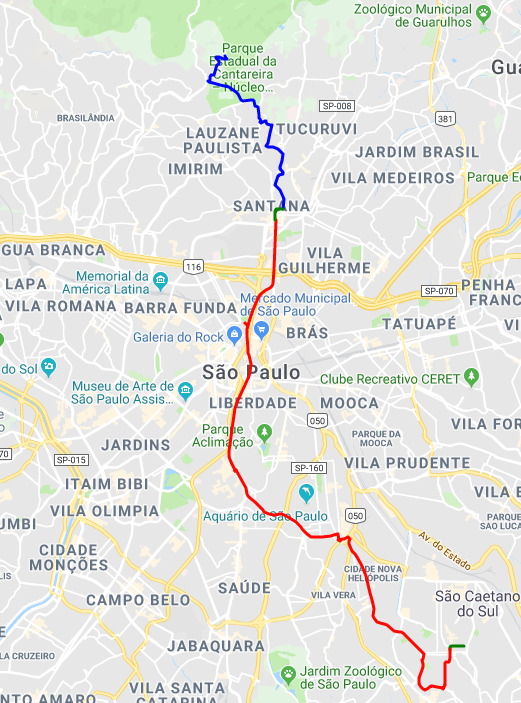}} 
    \subfigure[Híbrida 2.]{
        \includegraphics[scale=0.195]{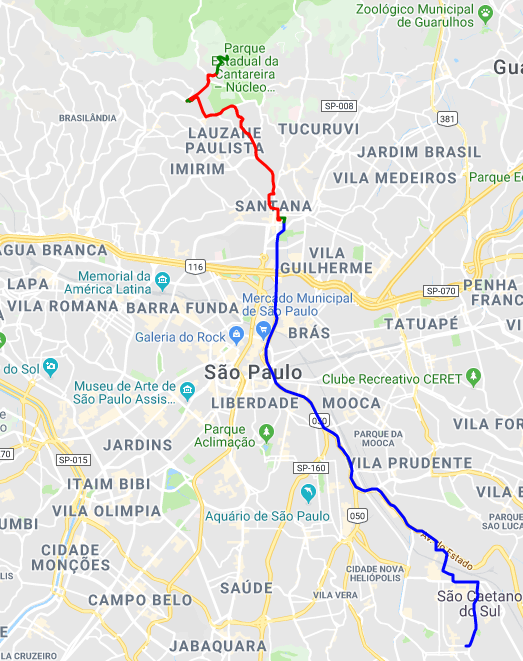}} 
    \caption{\textbf{Fluxo 2 --} Do bairro Olímpico, São Caetano do Sul, para o bairro Jardim Itatinga.}
    \label{fig:map_fluxo_2}
\end{figure}

\begin{figure}[!htp]
    \centering
    \subfigure[Transporte público.]{
        \includegraphics[scale=0.3]{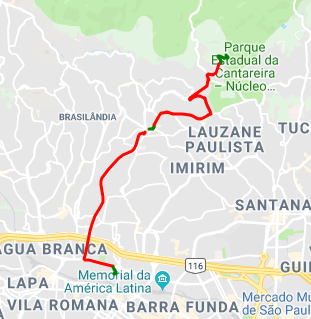}} 
    \subfigure[Uber.]{
        \includegraphics[scale=0.3]{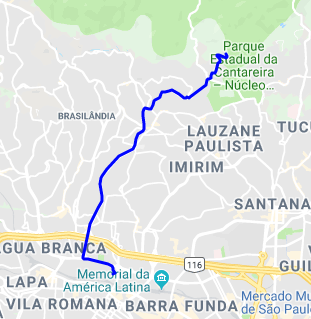}}
    \subfigure[Híbrida 1.]{
        \includegraphics[scale=0.3]{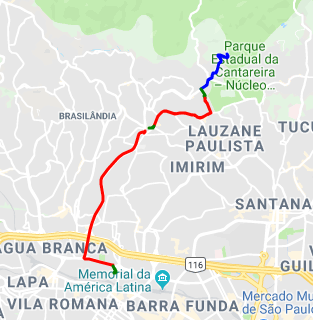}} 
    \subfigure[Híbrida 2.]{
        \includegraphics[scale=0.3]{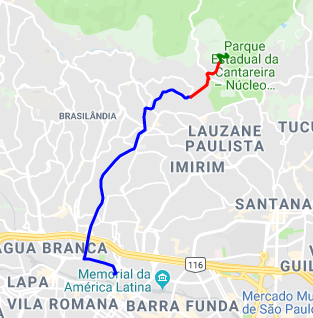}} 
    \caption{\textbf{Fluxo 3 --} Do bairro Jardim Itatinga para o bairro Jardim das Perdizes.}
    \label{fig:map_fluxo_3}
\end{figure}

\begin{figure}[!htp]
    \centering
    \subfigure[Transporte público.]{
        \includegraphics[scale=0.18]{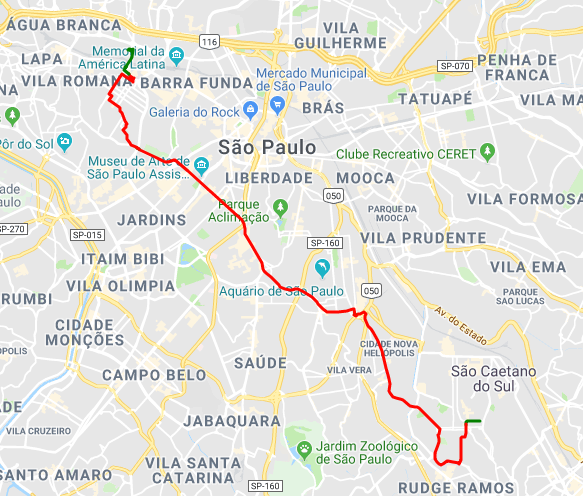}} 
    \subfigure[Uber.]{
        \includegraphics[scale=0.18]{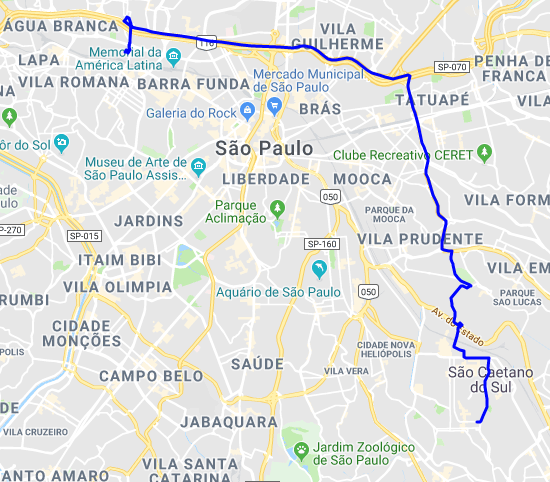}}
    \subfigure[Híbrida 1.]{
        \includegraphics[scale=0.18]{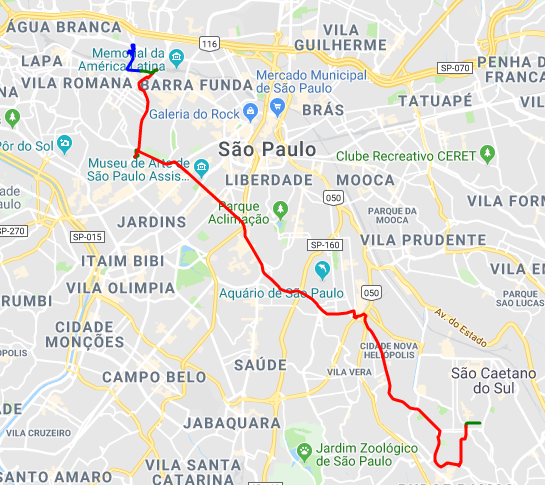}} 
    \subfigure[Híbrida 2.]{
        \includegraphics[scale=0.18]{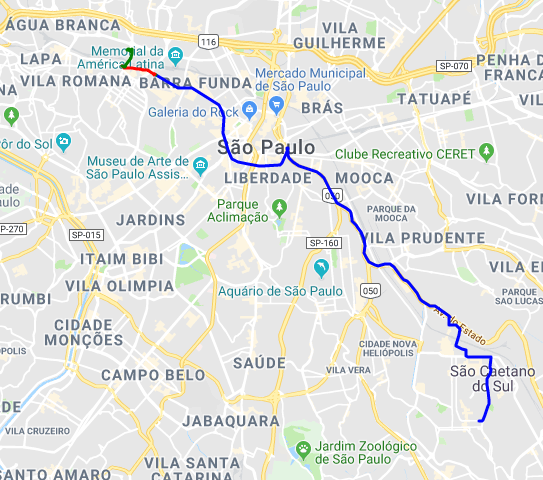}} 
    \caption{\textbf{Fluxo 4 --} Do bairro Olímpico, São Caetano do Sul, para o bairro Jardim das Perdizes.}
    \label{fig:map_fluxo_4}
\end{figure}

\begin{figure}[!htp]
    \centering
    \subfigure[Transporte público.]{
        \includegraphics[scale=0.18]{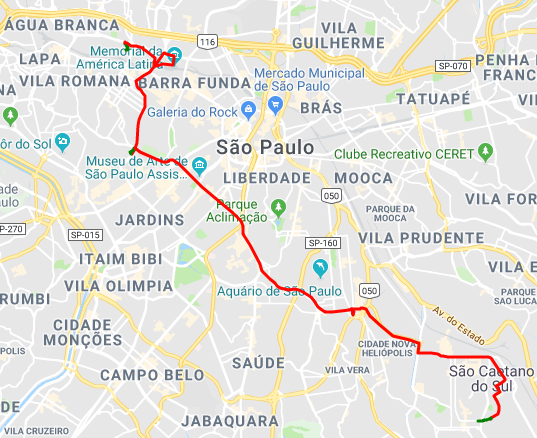}} 
    \subfigure[Uber.]{
        \includegraphics[scale=0.18]{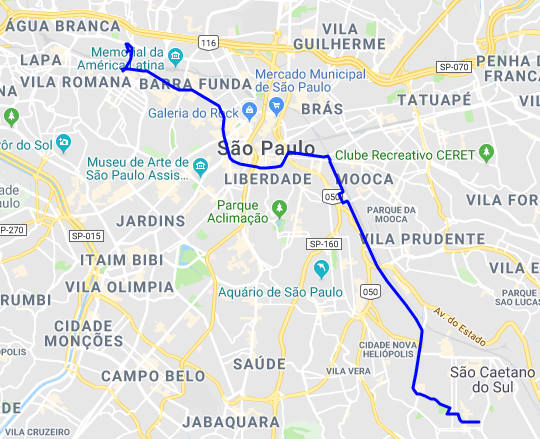}}
    \subfigure[Híbrida 1.]{
        \includegraphics[scale=0.18]{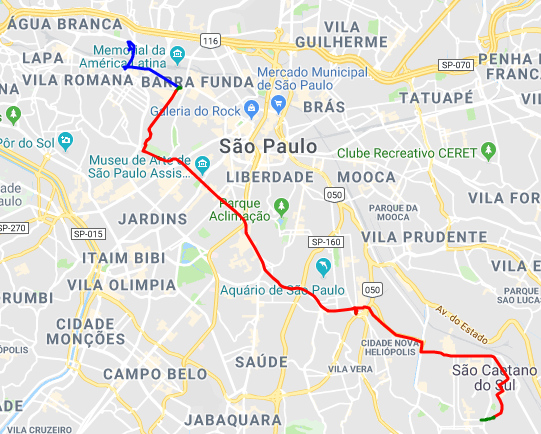}} 
    \subfigure[Híbrida 2.]{
        \includegraphics[scale=0.18]{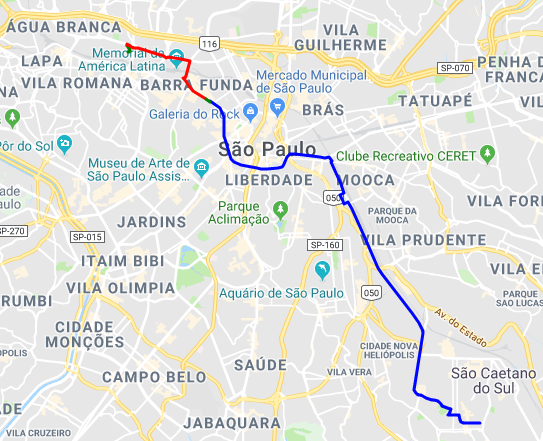}} 
    \caption{\textbf{Fluxo 5 --} Do bairro Jardim das Perdizes para o bairro Olímpico, São Caetano do Sul.}
    \label{fig:map_fluxo_5}
\end{figure}

\begin{figure}[!htp]
    \centering
    \subfigure[Transporte público.]{
        \includegraphics[scale=0.276]{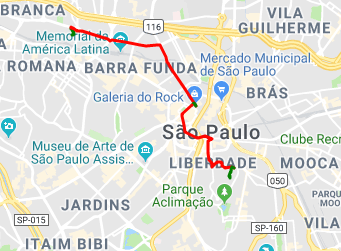}} 
    \subfigure[Uber.]{
        \includegraphics[scale=0.27]{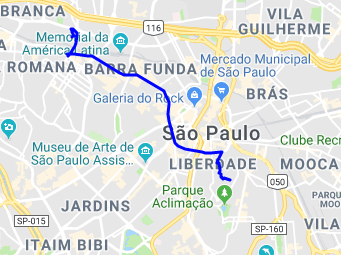}}
    \subfigure[Híbrida 1.]{
        \includegraphics[scale=0.292]{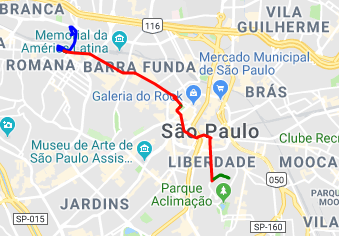}} 
    \subfigure[Híbrida 2.]{
        \includegraphics[scale=0.291]{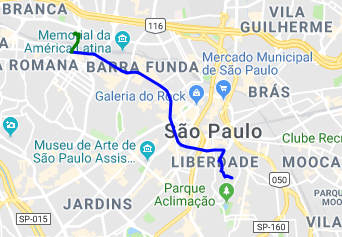}} 
    \caption{\textbf{Fluxo 6 --} Do bairro Jardim das Perdizes para o bairro Aclimação.}
    \label{fig:map_fluxo_6}
\end{figure}

\begin{figure}[!htp]
    \centering
    \subfigure[Transporte público.]{
        \includegraphics[scale=0.28]{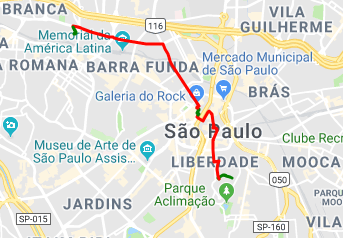}} 
    \subfigure[Uber.]{
        \includegraphics[scale=0.285]{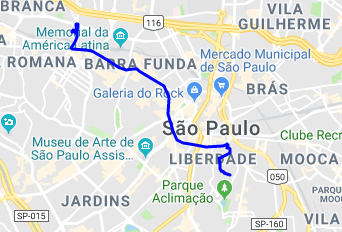}}
    \subfigure[Híbrida 1.]{
        \includegraphics[scale=0.285]{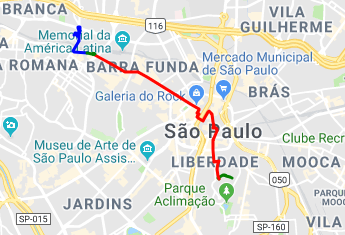}} 
    \subfigure[Híbrida 2.]{
        \includegraphics[scale=0.288]{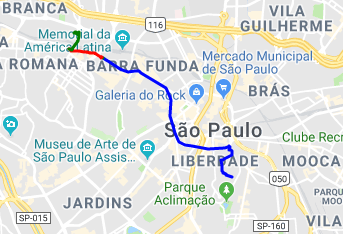}} 
    \caption{\textbf{Fluxo 7 --} Do bairro Aclimação para o bairro Jardim das Perdizes.}
    \label{fig:map_fluxo_7}
\end{figure}

\subsection{Análise Comparativa das Rotas} \label{subsec:estudodecaso:results}
\begin{figure}[!htp]
    \centering
    \subfigure[Distância.]{
        \includegraphics[scale=0.35]{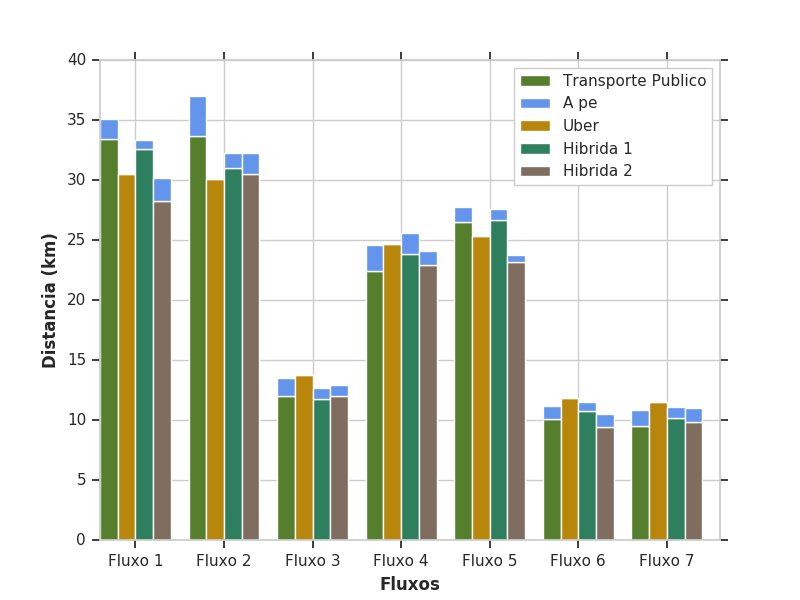}} 
    \subfigure[Tempo estimado de viagem.]{
        \includegraphics[scale=0.35]{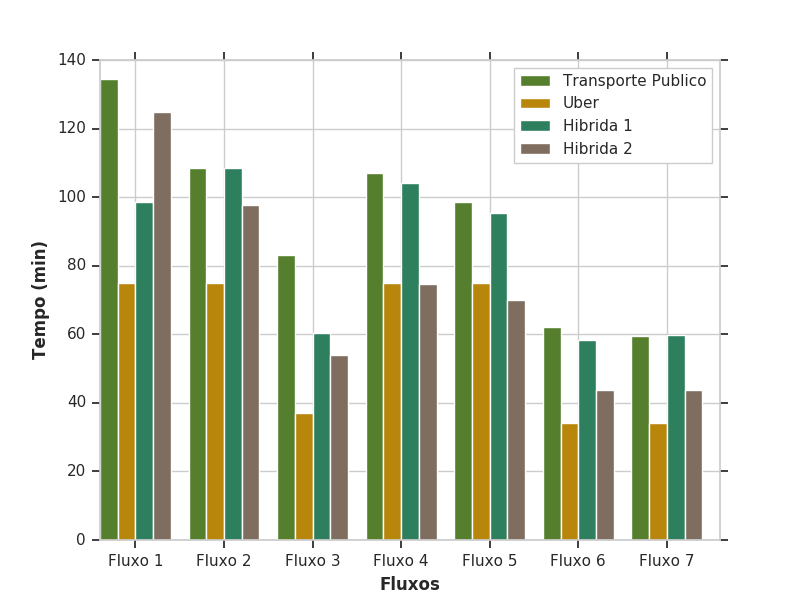}}
    \subfigure[Preço.]{
        \includegraphics[scale=0.35]{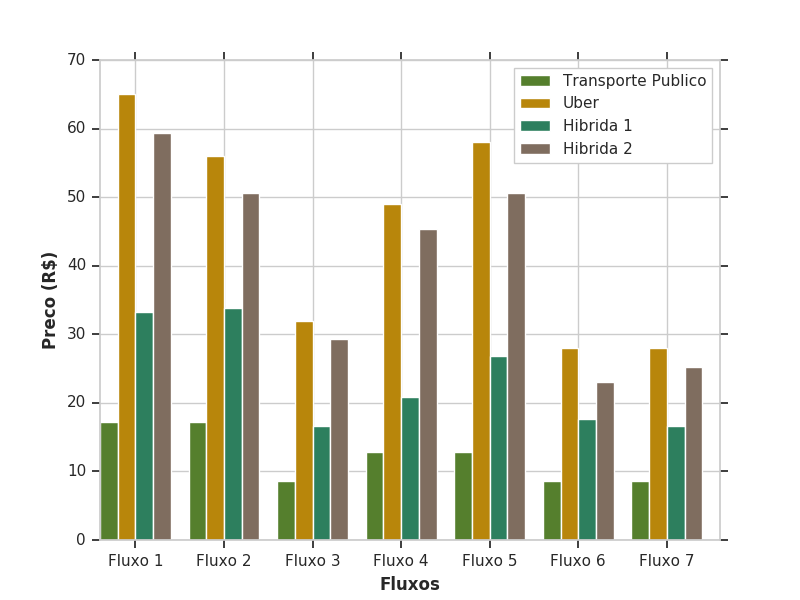}} 
    \subfigure[Tempo de espera.]{
        \includegraphics[scale=0.35]{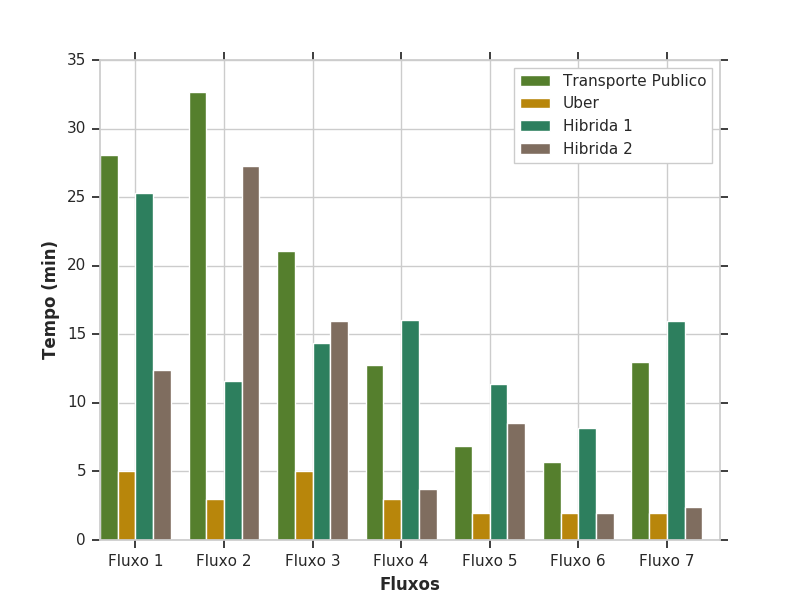}} 
    \caption{\textbf{Fluxo 3 --} Do bairro Jardim Itatinga para o bairro Jardim das Perdizes.}
    \label{fig:metricas_fluxos}
\end{figure}

Considerando as métricas de distância (total e percorrida a pé), preço, tempo estimado de viagem e tempo de espera, realizamos uma análise comparativa das rotas estudadas. Inicialmente, a Figura \ref{fig:metricas_fluxos} mostra os valores obtidos para essas métricas considerando as rotas apresentadas nas Figuras \ref{fig:map_fluxo_1}--\ref{fig:map_fluxo_7}. Com relação a distância (Figura \ref{fig:metricas_fluxos}(a)), os valores obtidos pelas rotas multimodais Híbrida 1 e Híbrida 2 são muito semelhantes as rotas Transporte Público e Uber, respectivamente, para todos os fluxos. Este resultado era esperado, uma vez que ambas as rotas multimodais são compostas em sua maioria por um desses dois tipos de meio de transporte. Diferentemente, as outras três métricas apresentam variações significativas na maioria dos fluxos quando são utilizadas rotas multimodais em relação as rotas tradicionais. Por exemplo, o tempo estimado de viagem (Figura \ref{fig:metricas_fluxos}(b)) das rotas Híbrida 2 aumentam consideravelmente em relação as rotas Uber para os fluxos 1, 2, 3, 6 e 7, refletindo o impacto em também ser utilizado o transporte público em conjunto com o Uber, ainda que o modo de transporte Uber seja majoritário nas rotas Híbrida 2. Similarmente, o tempo estimado de viagem das rotas Híbrida 1 diminuem significativamente para os fluxos 1 e 3 em relação as rotas Transporte Público, e, apesar de manter-se similar para os demais fluxos, note que é reduzida a distância percorrida a pé em todos os fluxos quando comparado esses dois tipos de rota. Em relação ao preço (Figura \ref{fig:metricas_fluxos}(c)) e ao tempo de espera (Figura \ref{fig:metricas_fluxos}(d)), é possível observar um relacionamento inverso entre essas métricas, i.e., quanto maior o preço a ser pago, menor é o tempo de espera e vice-versa. É importante ressaltar que foi considerado que o transporte público tem preço R\$ 4,30, que é referente a taxa cobrada por trecho na cidade de São Paulo, podendo ser diferente para outros municípios. É possível observar que houveram duas exceções a esse relacionamento, os fluxos 2 e 3, onde a rota Híbrida 1 tem o preço e tempo de espera inferior ao da rota Híbrida 2 em ambos os fluxos. Situações como essa se deve ao momento em que os valores para as métricas foram calculados, onde pode haver uma boa sincronia entre os serviços sem requerer que o usuário aguarde muito entre um meio de transporte e outro. Para evitar casos pontuais como esse, pode-se realizar um experimento empírico, executando o algoritmo \ref{alg:compute-hybrid-routes} em momentos distintos do dia (e.g., manhã, tarde e noite) ao longo de vários dias (e.g., um mês). Assim, é possível obter valores médios para as métricas sem que eventos esporádicos tenham impactos significativos nesses valores.

\begin{figure}[!htp]
    \centering
    \subfigure[Preço]{
        \includegraphics[scale=0.35]{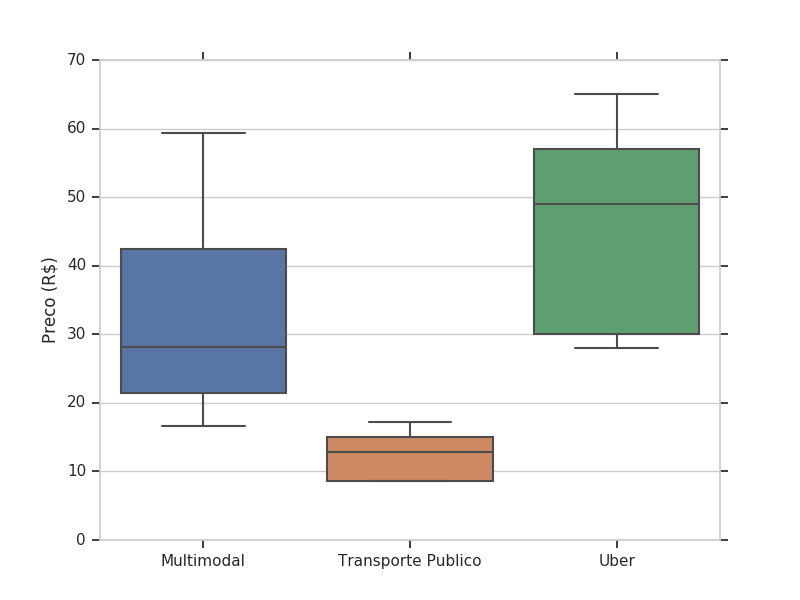}}
    \subfigure[Tempo estimado de viagem.]{
        \includegraphics[scale=0.35]{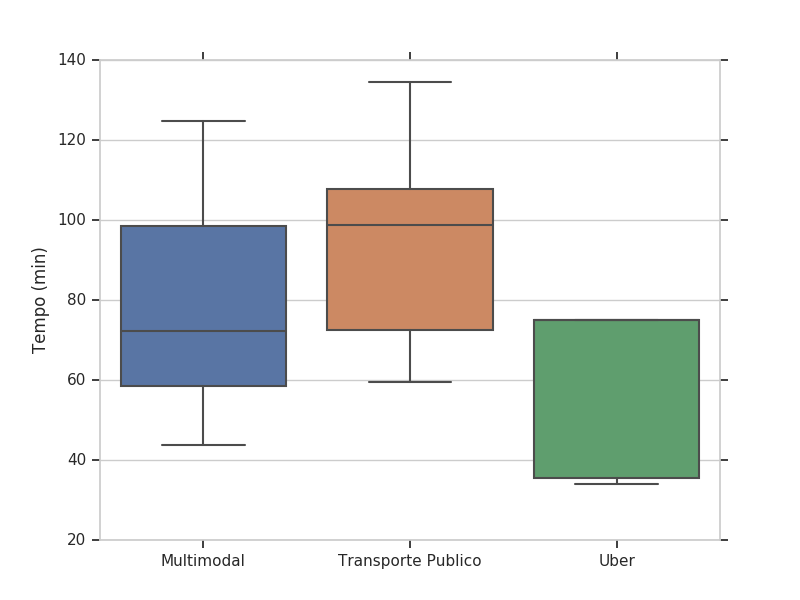}}
    \subfigure[Tempo de espera.]{
        \includegraphics[scale=0.35]{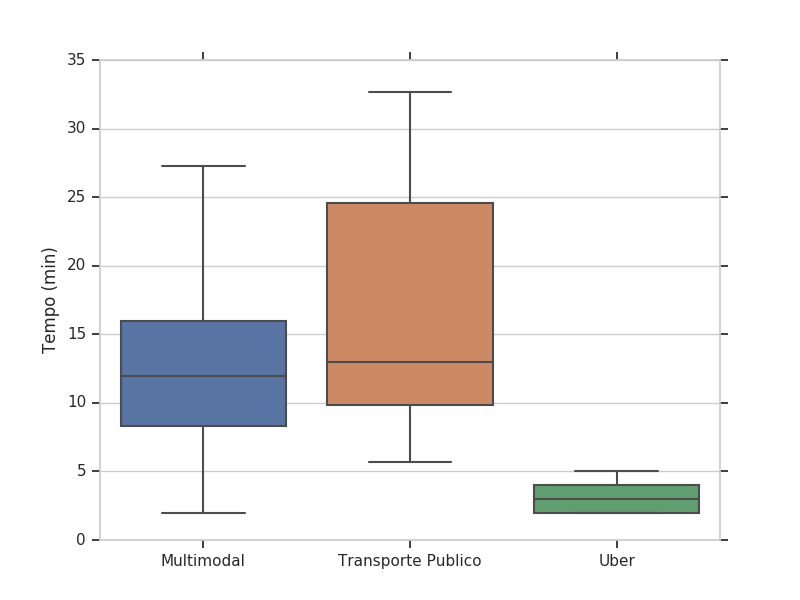}}
    \subfigure[Distância.]{
        \includegraphics[scale=0.35]{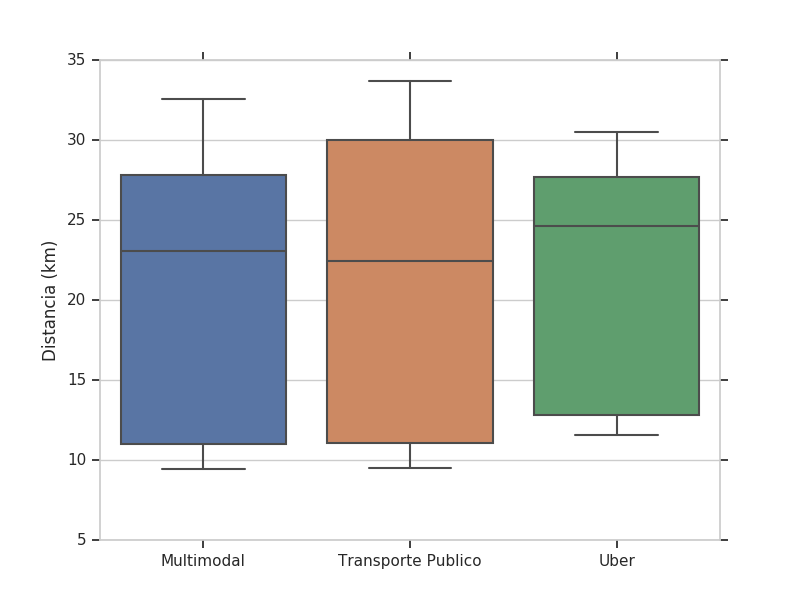}}
    \subfigure[Distância percorrida a pé.]{
        \includegraphics[scale=0.35]{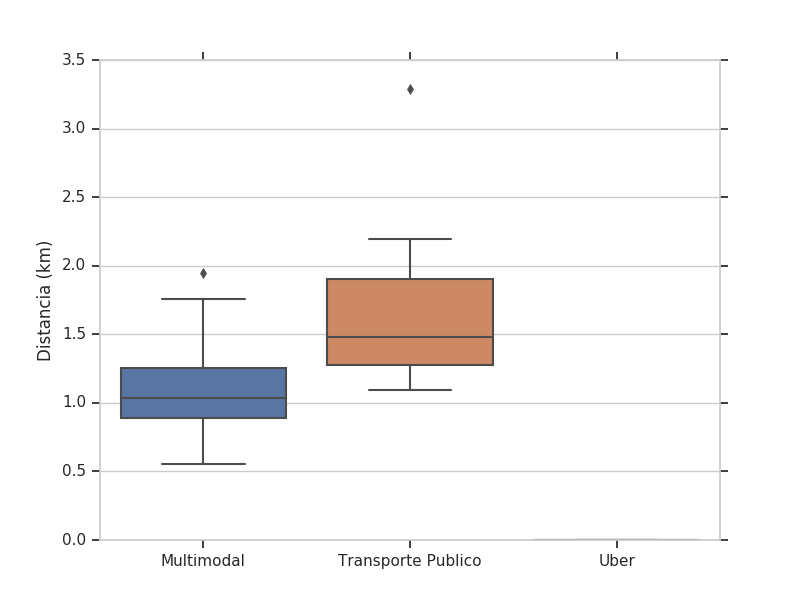}}
        
    \caption{Desempenho médio das rotas multimodais e tradicionais, considerando os fluxos de 1 à 7.}
    \label{fig:estudodecaso:impacto}
\end{figure}

Também é avaliado o impacto médio do uso de rotas multimodais (Híbrida 1 e 2) em comparação com as rotas tradicionais (Transporte Público e Uber) em todos os fluxos estudados, como mostra a Figura \ref{fig:estudodecaso:impacto}. Como podemos observar, com exceção ao preço, as rotas multimodais são mais viáveis do que as rotas com Transporte Público em todas as demais métricas, onde as principais vantagens são a redução do tempo de viagem médio (cerca de 23 minutos menor) e a redução da distância percorrida a pé. Apesar das rotas com Uber ter a clara vantagem de não ser necessário realizar deslocamentos a pé e, ainda, ter baixo tempos de espera e de viagem quando comparado com as demais, o preço pode ser um fator crucial na escolha de muitos usuários, tornando esse tipo de rota inviável já que, na média, são cerca de 2 vezes mais caras que as rotas multimodais. Desta forma, pode-se concluir que as rotas multimodais sugeridas pela aplicação alcançam um bom desempenho, sendo mais uma opção para o usuário.

O trabalho \textit{Hybrid Context-Aware Multimodal Routing} \cite{RodriguesItsc2018} realiza experimentos similares aos realizados aqui, onde o cenário utilizado é a cidade de Nova Iorque, NY, EUA, e propõem uma função de otimização para calcular a experiência do usuário de acordo com as mesmas métricas apresentadas nesta seção, com o intuito de apoiar no processo de tomada de decisão. Aqui, optamos em apresentar e analisar dois exemplos de rotas multimodais além das rotas tradicionais. Contudo, o único critério para seleção das rotas multimodais foi a composição delas, sendo os dois extremos possíveis (i.e., uma sendo a maior parte com o meio de transporte Uber e a outra com Transporte Público), mas nenhuma das métricas foi utilizada individualmente (e.g., a rota multimodal de menor preço ou tempo de viagem) ou em conjunto (e.g., a rota de menor preço e com menor distância percorrida a pé) para seleção dessas rotas. Caso o leitor tenha a intenção de selecionar as ``melhores'' rotas multimodais para serem comparadas com as rotas tradicionais, métodos de otimização multiobjetivos devem ser explorados e incorporados à aplicação.



\section{Conclusão}
\label{sec:conclusao}

Devido ao expressivo volume de dados gerados pelos espaços urbanos, bem como nos avanços da tecnologia da informação, estamos diante de uma oportunidade sem precedentes para estudo da Computação Urbana. Diante disso, este minicurso introduziu os fundamentos da Computação Urbana e apresentou os meios necessários para desenvolver uma aplicação nessa área. Para isso, explorou os conceitos que englobam uma visão geral dos trabalhos existentes, os desafios e as perspectivas futuras nessa área em questão. Isso foi útil para apresentar as tendências e ferramentas tipicamente empregadas para o desenvolvimento das aplicações com base na Computação Urbana. 

\bibliographystyle{sbc}
\bibliography{refs}

\begin{thebibliography}{}

\bibitem[Akabane et~al. 2018a]{akabane2018imob}
Akabane, A.~T., Immich, R., Madeira, E.~R., and Villas, L.~A. (2018a).
\newblock imob: An intelligent urban mobility management system based on
  vehicular social networks.
\newblock In {\em 2018 IEEE Vehicular Networking Conference (VNC)}, pages 1--8.
  IEEE.

\bibitem[Akabane et~al. 2018b]{akabane2018trusted}
Akabane, A.~T., Immich, R., Pazzi, R.~W., Madeira, E.~R., and Villas, L.~A.
  (2018b).
\newblock Trusted: A distributed system for information management and
  knowledge distribution in vanets.
\newblock In {\em 2018 IEEE Symposium on Computers and Communications (ISCC)},
  pages 1--6. IEEE.

\bibitem[Barnaghi et~al. 2012]{Barnaghi2012}
Barnaghi, P., Wang, W., Henson, C., and Taylor, K. (2012).
\newblock Semantics for the internet of things: Early progress and back to the
  future.
\newblock {\em Int. J. Semant. Web Inf. Syst.}, 8(1):1--21.

\bibitem[Bellavista and Zanni 2017]{Bellavista2017}
Bellavista, P. and Zanni, A. (2017).
\newblock Feasibility of fog computing deployment based on docker
  containerization over raspberrypi.
\newblock In {\em Proceedings of the 18th International Conference on
  Distributed Computing and Networking}, ICDCN '17, pages 16:1--16:10, New
  York, NY, USA. ACM.

\bibitem[Birari and Iyer 2005]{birari2005mitigating}
Birari, S.~M. and Iyer, S. (2005).
\newblock Mitigating the reader collision problem in rfid networks with mobile
  readers.
\newblock In {\em 2005 13th IEEE International Conference on Networks Jointly
  held with the 2005 IEEE 7th Malaysia International Conf on Communic},
  volume~1, pages 6--pp. IEEE.

\bibitem[Bittencourt et~al. 2018]{BITTENCOURT2018}
Bittencourt, L., Immich, R., Sakellariou, R., Fonseca, N., Madeira, E., Curado,
  M., Villas, L., da~Silva, L., Lee, C., and Rana, O. (2018).
\newblock The internet of things, fog and cloud continuum: Integration and
  challenges.
\newblock {\em Internet of Things}.

\bibitem[Bittencourt et~al. 2015]{7424534}
Bittencourt, L.~F., Lopes, M.~M., Petri, I., and Rana, O.~F. (2015).
\newblock Towards virtual machine migration in fog computing.
\newblock In {\em 2015 10th International Conference on P2P, Parallel, Grid,
  Cloud and Internet Computing (3PGCIC)}, pages 1--8.

\bibitem[Byers 2017]{Byers2017}
Byers, C.~C. (2017).
\newblock Architectural imperatives for fog computing: Use cases, requirements,
  and architectural techniques for fog-enabled iot networks.
\newblock {\em IEEE Communications Magazine}, 55(8):14--20.

\bibitem[{Celes} et~al. 2017]{celes2017}
{Celes}, C., {Silva}, F.~A., {Boukerche}, A., d.~C.~{Andrade}, R.~M., and
  {Loureiro}, A. A.~F. (2017).
\newblock Improving vanet simulation with calibrated vehicular mobility traces.
\newblock {\em IEEE Transactions on Mobile Computing}, 16(12):3376--3389.

\bibitem[Chen et~al. 2016]{chen2016dynamic}
Chen, N., Chen, Y., You, Y., Ling, H., Liang, P., and Zimmermann, R. (2016).
\newblock Dynamic urban surveillance video stream processing using fog
  computing.
\newblock In {\em 2016 IEEE second international conference on multimedia big
  data (BigMM)}, pages 105--112. IEEE.

\bibitem[Confais et~al. 2017]{Confais2017}
Confais, B., Lebre, A., and Parrein, B. (2017).
\newblock {\em Performance Analysis of Object Store Systems in a Fog and Edge
  Computing Infrastructure}, pages 40--79.
\newblock Springer Berlin Heidelberg, Berlin, Heidelberg.

\bibitem[Cunha et~al. 2017]{cunha2017sistemas}
Cunha, F., Maia, G., Celes, C., Guidoni, D., de~Souza, F., Ramos, H., and
  Villas, L. (2017).
\newblock Sistemas de transporte inteligentes: Conceitos, aplica{\c{c}}ões
  desafios.
\newblock {\em Livro de Minicursos do Simp{\'o}sio Brasileiro de Redes de
  Computadores e Sistemas Distribuídos (SBRC’17)}.

\bibitem[Curado et~al. 2018]{Curado2019}
Curado, M., Madeira, H., da~Cunha, P.~R., Cabral, B., Abreu, D.~P., Barata, J.,
  Roque, L., and Immich, R. (2018).
\newblock {\em Internet of Things}, pages 381--401.
\newblock Springer International Publishing.

\bibitem[De~Souza et~al. 2017]{de2017traffic}
De~Souza, A.~M., Brennand, C.~A., Yokoyama, R.~S., Donato, E.~A., Madeira,
  E.~R., and Villas, L.~A. (2017).
\newblock Traffic management systems: A classification, review, challenges, and
  future perspectives.
\newblock {\em International Journal of Distributed Sensor Networks},
  13(4):1550147716683612.

\bibitem[de~Souza et~al. 2018]{DeSouza2018}
de~Souza, A.~M., Pedrosa, L. L.~C., Botega, L.~C., and Villas, L. (2018).
\newblock {Itssafe: An Intelligent Transportation System for Improving Safety
  and Traffic Efficiency}.
\newblock In {\em 2018 IEEE 87th Vehicular Technology Conference (VTC Spring)},
  pages 1--7. IEEE.

\bibitem[Dib et~al. 2015]{dib2015memetic}
Dib, O., Manier, M.-A., and Caminada, A. (2015).
\newblock Memetic algorithm for computing shortest paths in multimodal
  transportation networks.
\newblock {\em Transportation Research Procedia}, 10:745--755.

\bibitem[Duque et~al. 2015]{duque2015exact}
Duque, D., Lozano, L., and Medaglia, A.~L. (2015).
\newblock An exact method for the biobjective shortest path problem for
  large-scale road networks.
\newblock {\em European Journal of Operational Research}, 242(3):788--797.

\bibitem[{Estrin} et~al. 2010]{Estrin2010}
{Estrin}, D., {Chandy}, K.~M., {Young}, R.~M., {Smarr}, L., {Odlyzko}, A.,
  {Clark}, D., {Reding}, V., {Ishida}, T., {Sharma}, S., {Cerf}, V.~G.,
  {HÃlzle}, U., {Barroso}, L.~A., {Mulligan}, G., {Hooke}, A., and {Elliott},
  C. (2010).
\newblock Participatory sensing: applications and architecture [internet
  predictions].
\newblock {\em IEEE Internet Computing}, 14(1):12--42.

\bibitem[Golestan et~al. 2015]{golestan2015localization}
Golestan, K., Sattar, F., Karray, F., Kamel, M., and Seifzadeh, S. (2015).
\newblock Localization in vehicular ad hoc networks using data fusion and v2v
  communication.
\newblock {\em Computer Communications}, 71:61--72.

\bibitem[Group et~al. 2017]{openfog_2017}
Group, O. C. A.~W. et~al. (2017).
\newblock Openfog reference architecture for fog computing.
\newblock {\em OPFRA001}, 20817:162.

\bibitem[Guo et~al. 2014]{Guo:TTT:2014}
Guo, C., Jensen, C.~S., and Yang, B. (2014).
\newblock Towards total traffic awareness.
\newblock {\em SIGMOD Record}, 43(3):18--23.

\bibitem[Heintz et~al. 2016]{Heintz2016}
Heintz, B., Chandra, A., Sitaraman, R.~K., and Weissman, J. (2016).
\newblock End-to-end optimization for geo-distributed mapreduce.
\newblock {\em IEEE Transactions on Cloud Computing}, 4(3):293--306.

\bibitem[Hung et~al. 2015]{Hung2015}
Hung, C.-C., Golubchik, L., and Yu, M. (2015).
\newblock Scheduling jobs across geo-distributed datacenters.
\newblock In {\em Proceedings of the Sixth ACM Symposium on Cloud Computing},
  SoCC '15, pages 111--124, New York, NY, USA. ACM.

\bibitem[Idri et~al. 2017]{idri2017new}
Idri, A., Oukarfi, M., Boulmakoul, A., Zeitouni, K., and Masri, A. (2017).
\newblock A new time-dependent shortest path algorithm for multimodal
  transportation network.
\newblock {\em Procedia Computer Science}, 109:692--697.

\bibitem[{Immich} et~al. 2015]{7179389}
{Immich}, R., {Cerqueira}, E., and {Curado}, M. (2015).
\newblock Adaptive qoe-driven video transmission over vehicular ad-hoc
  networks.
\newblock In {\em 2015 IEEE Conference on Computer Communications Workshops
  (INFOCOM WKSHPS)}, pages 227--232.

\bibitem[Immich et~al. 2018]{Immich2018}
Immich, R., Cerqueira, E., and Curado, M. (2018).
\newblock Efficient high-resolution video delivery over vanets.
\newblock {\em Wireless Networks}.

\bibitem[Issarny et~al. 2018]{issarny2018service}
Issarny, V., Bouloukakis, G., Georgantas, N., Sailhan, F., and Texier, G.
  (2018).
\newblock When service-oriented computing meets the iot: A use case in the
  context of urban mobile crowdsensing.
\newblock In {\em European Conference on Service-Oriented and Cloud Computing},
  pages 1--16. Springer.

\bibitem[{Ji} et~al. 2017]{ji2017}
{Ji}, S., {Mittal}, P., and {Beyah}, R. (2017).
\newblock Graph data anonymization, de-anonymization attacks, and
  de-anonymizability quantification: A survey.
\newblock {\em IEEE Communications Surveys Tutorials}, 19(2):1305--1326.

\bibitem[Kim 2015]{foursquareRec}
Kim, S. (2015).
\newblock {\em How Foursquare and Other Apps Guess What You Want to Eat}.
\newblock Eater.com.
\newblock https://goo.gl/icFBrH.

\bibitem[Krasnogor and Smith 2005]{krasnogor2005tutorial}
Krasnogor, N. and Smith, J. (2005).
\newblock A tutorial for competent memetic algorithms: model, taxonomy, and
  design issues.
\newblock {\em IEEE Transactions on Evolutionary Computation}, 9(5):474--488.

\bibitem[Meneguette et~al. 2016a]{meneguette2016solution}
Meneguette, R., Fillho, G., Bittencourt, L., Ueyama, J., and Villas, L.
  (2016a).
\newblock A solution for detection and control for congested roads using
  vehicular networks.
\newblock {\em IEEE Latin America Transactions}, 14(4):1849--1855.

\bibitem[Meneguette et~al. 2016b]{meneguette2016increasing}
Meneguette, R.~I., Geraldo~Filho, P., Guidoni, D.~L., Pessin, G., Villas,
  L.~A., and Ueyama, J. (2016b).
\newblock Increasing intelligence in inter-vehicle communications to reduce
  traffic congestions: experiments in urban and highway environments.
\newblock {\em PLoS one}, 11(8):e0159110.

\bibitem[Moura et~al. 2018]{Moura2018}
Moura, D.~L., Cabral, R.~S., Sales, T., and Aquino, A.~L. (2018).
\newblock An evolutionary algorithm for roadside unit deployment with
  betweenness centrality preprocessing.
\newblock {\em Future Generation Computer Systems}, 88:776 -- 784.

\bibitem[Mueller et~al. 2017]{Mueller2017}
Mueller, W., Silva, T.~H., Almeida, J.~M., and Loureiro, A.~A. (2017).
\newblock Gender matters! analyzing global cultural gender preferences for
  venues using social sensing.
\newblock {\em EPJ Data Science}, 6(1):5.

\bibitem[Naboulsi and Fiore 2013]{Naboulsi2013}
Naboulsi, D. and Fiore, M. (2013).
\newblock On the instantaneous topology of a large-scale urban vehicular
  network: the cologne case.
\newblock In {\em Proceedings of the fourteenth ACM international symposium on
  Mobile ad hoc networking and computing}, pages 167--176. ACM.

\bibitem[Nascimento et~al. 2018]{nascimento2018integrated}
Nascimento, P., Kimura, B., Guidoni, D., and Villas, L. (2018).
\newblock An integrated dead reckoning with cooperative positioning solution to
  assist gps nlos using vehicular communications.
\newblock {\em Sensors}, 18(9):2895.

\bibitem[Ning et~al. 2019]{ning2019vehicular}
Ning, Z., Huang, J., and Wang, X. (2019).
\newblock Vehicular fog computing: Enabling real-time traffic management for
  smart cities.
\newblock {\em IEEE Wireless Communications}, 26(1):87--93.

\bibitem[Ribeiro et~al. 2014]{Ribeiro2014}
Ribeiro, A. I. J.~a.~T., Silva, T.~H., Duarte-Figueiredo, F., and Loureiro,
  A.~A. (2014).
\newblock Studying traffic conditions by analyzing foursquare and instagram
  data.
\newblock In {\em Proceedings of the 11th ACM Symposium on Performance
  Evaluation of Wireless Ad Hoc, Sensor, \&\#38; Ubiquitous Networks}, PE-WASUN
  '14, pages 17--24, New York, NY, USA. ACM.

\bibitem[Riveiro et~al. 2017]{riveiro2017anomaly}
Riveiro, M., Lebram, M., and Elmer, M. (2017).
\newblock Anomaly detection for road traffic: A visual analytics framework.
\newblock {\em IEEE Transactions on Intelligent Transportation Systems},
  18(8):2260--2270.

\bibitem[{Rocha Filho} et~al. 2018a]{geraldo2018sistema}
{Rocha Filho}, G.~P., Neto, J. R.~T., Valejo, A., Meneguette, R.~I., Villas,
  L.~A., and Ueyama, J. (2018a).
\newblock Um sistema de controle neuro-fog para infraestruturas residenciais
  via objetos inteligentes.
\newblock In {\em Anais do XXXVI Simp{\'o}sio Brasileiro de Redes de
  Computadores e Sistemas Distribu{\'\i}dos}. SBC.

\bibitem[{Rocha Filho} et~al. 2013]{geraldo}
{Rocha Filho}, G.~P., Ueyama, J., Villas, L., and Seraphini, A. P.~S. (2013).
\newblock Nodepm: Um sistema de monitoramento remoto do consumo de energia
  eletrica via redes de sensores sem fio.
\newblock In sociedade Brasileira~de Computacao~(SBC), editor, {\em Simposio
  Brasileiro de Redes de Computadores e Sistemas Distribuidos (SBRC)},
  volume~31, pages 17--30.

\bibitem[{Rocha Filho} et~al. 2018b]{PRFILHO201854}
{Rocha Filho}, G.~P., Villas, L.~A., Freitas, H., Valejo, A., Guidoni, D.~L.,
  and Ueyama, J. (2018b).
\newblock Residi: Towards a smarter smart home system for decision-making using
  wireless sensors and actuators.
\newblock {\em Computer Networks}, 135:54 -- 69.

\bibitem[{Rocha Filho} et~al. 2019]{FILHO2019153}
{Rocha Filho}, G.~P., Villas, L.~A., Gonçalves, V.~P., Pessin, G., Loureiro,
  A.~A., and Ueyama, J. (2019).
\newblock Energy-efficient smart home systems: Infrastructure and
  decision-making process.
\newblock {\em Internet of Things}, 5:153 -- 167.

\bibitem[{Rocha Filho} et~al. 2018c]{8291472}
{Rocha Filho}, G.~P., {Yukio Mano}, L., {Demetrius Baria Valejo}, A.,
  {Aparecido Villas}, L., and {Ueyama}, J. (2018c).
\newblock A low-cost smart home automation to enhance decision-making based on
  fog computing and computational intelligence.
\newblock {\em IEEE Latin America Transactions}, 16(1):186--191.

\bibitem[Rodrigues et~al. 2018]{RODRIGUES2018111}
Rodrigues, D.~O., Boukerche, A., Silva, T.~H., Loureiro, A.~A., and Villas,
  L.~A. (2018).
\newblock Combining taxi and social media data to explore urban mobility
  issues.
\newblock {\em Computer Communications}, 132:111 -- 125.

\bibitem[{Rodrigues} et~al. 2018]{RodriguesItsc2018}
{Rodrigues}, D.~O., {Fernandes}, J.~T., {Curado}, M., and {Villas}, L.~A.
  (2018).
\newblock Hybrid context-aware multimodal routing.
\newblock In {\em 2018 21st International Conference on Intelligent
  Transportation Systems (ITSC)}, pages 2250--2255.

\bibitem[{S. Gama} et~al. 2018]{Gama2018}
{S. Gama}, E., {Immich}, R., and {F. Bittencourt}, L. (2018).
\newblock Towards a multi-tier fog/cloud architecture for video streaming.
\newblock In {\em 2018 IEEE/ACM International Conference on Utility and Cloud
  Computing Companion (UCC Companion)}, pages 13--14.

\bibitem[Sakr et~al. 2011]{Sakr2011}
Sakr, S., Liu, A., Batista, D.~M., and Alomari, M. (2011).
\newblock A survey of large scale data management approaches in cloud
  environments.
\newblock {\em IEEE Communications Surveys Tutorials}, 13(3):311--336.

\bibitem[Santana et~al. 2018]{santana2018software}
Santana, E. F.~Z., Chaves, A.~P., Gerosa, M.~A., Kon, F., and Milojicic, D.~S.
  (2018).
\newblock Software platforms for smart cities: Concepts, requirements,
  challenges, and a unified reference architecture.
\newblock {\em ACM Computing Surveys (CSUR)}, 50(6):78.

\bibitem[Santos et~al. 2018]{santosWI2018}
Santos, F.~A., Silva, T.~H., , Loureiro, A. A.~F., and Villas, L.~A. (2018).
\newblock {Uncovering the Perception of Urban Outdoor Areas Expressed in Social
  Media}.
\newblock In {\em Proc. of IEEE ACM International Conference on Web
  Intelligence (WI)}, Santiago, Chile.

\bibitem[Santos et~al. 2017]{santos2017towards}
Santos, F.~A., Silva, T.~H., Braun, T., Loureiro, A.~A., and Villas, L.~A.
  (2017).
\newblock Towards a sustainable people-centric sensing.
\newblock In {\em 2017 IEEE International Conference on Communications (ICC)},
  pages 1--6. IEEE.

\bibitem[Sedeno-Noda and Raith 2015]{sedeno2015dijkstra}
Sedeno-Noda, A. and Raith, A. (2015).
\newblock A dijkstra-like method computing all extreme supported non-dominated
  solutions of the biobjective shortest path problem.
\newblock {\em Computers \& Operations Research}, 57:83--94.

\bibitem[Shi et~al. 2016]{shi2016edge}
Shi, W., Cao, J., Zhang, Q., Li, Y., and Xu, L. (2016).
\newblock Edge computing: Vision and challenges.
\newblock {\em IEEE Internet of Things Journal}, 3(5):637--646.

\bibitem[Silva 2016]{silvaUserUrbSensing}
Silva, T~H., d. M. P. N. J. I. A. d. R. C. M. V. d. C. F. F. A. A. J. L.~A.
  (2016).
\newblock Users in the urban sensing process: Challenges and research
  opportunities.
\newblock In {\em Pervasive Computing: Next Generation Platforms for
  Intelligent Data Collection}, pages 45--95. Academic Press.

\bibitem[Silva et~al. 2018]{Silva2018}
Silva, T., Viana, A., Benevenuto, F., Villas, L., Salles, J., Loureiro, A., and
  Quercia, D. (2018).
\newblock {Urban Computing Leveraging Location-Based Social Network Data: a
  Survey}.
\newblock {\em ACM Computing Surveys}.

\bibitem[Silva et~al. 2014]{Silva2014toit}
Silva, T.~H., Vaz~de Melo, P. O.~S., Almeida, J.~M., Salles, J., and Loureiro,
  A. A.~F. (2014).
\newblock Revealing the city that we cannot see.
\newblock {\em ACM Trans. Internet Technol.}, 14(4):26:1--26:23.

\bibitem[Silva et~al. 2013]{silvaSocInfo2013}
Silva, T.~H., Vaz~de Melo, P. O.~S., Viana, A., Almeida, J.~M., Salles, J., and
  Loureiro, A. A.~F. (2013).
\newblock {Traffic Condition is more than Colored Lines on a Map:
  Characterization of Waze Alerts}.
\newblock In {\em Proc.\ of the International Conference on Social Informatics
  (SocInfo'13)}, Kyoto, Japan.

\bibitem[Song et~al. 2010]{song2010limits}
Song, C., Qu, Z., Blumm, N., and Barab{\'a}si, A.-L. (2010).
\newblock Limits of predictability in human mobility.
\newblock {\em Science}, 327(5968):1018--1021.

\bibitem[Song et~al. 2013]{song2013modeling}
Song, X., Zhang, Q., Sekimoto, Y., Horanont, T., Ueyama, S., and Shibasaki, R.
  (2013).
\newblock Modeling and probabilistic reasoning of population evacuation during
  large-scale disaster.
\newblock In {\em Proceedings of the 19th ACM SIGKDD international conference
  on Knowledge discovery and data mining}, pages 1231--1239. ACM.

\bibitem[SteadieSeifi et~al. 2014]{steadieseifi2014multimodal}
SteadieSeifi, M., Dellaert, N.~P., Nuijten, W., Van~Woensel, T., and Raoufi, R.
  (2014).
\newblock Multimodal freight transportation planning: A literature review.
\newblock {\em European journal of operational research}, 233(1):1--15.

\bibitem[Taleb et~al. 2017a]{Taleb2017}
Taleb, T., Dutta, S., Ksentini, A., Iqbal, M., and Flinck, H. (2017a).
\newblock Mobile edge computing potential in making cities smarter.
\newblock {\em Comm. Mag.}, 55(3):38--43.

\bibitem[Taleb et~al. 2017b]{Taleb2017b}
Taleb, T., Ksentini, A., and Frangoudis, P. (2017b).
\newblock Follow-me cloud: When cloud services follow mobile users.
\newblock {\em IEEE Transactions on Cloud Computing}, pages 1--1.

\bibitem[Traynor and Curran 2012]{traynor2012location}
Traynor, D. and Curran, K. (2012).
\newblock Location-based social networks.
\newblock {\em From Government to E-Governance: Public Administration in the
  Digital Age}, page 243.

\bibitem[United~Nations and Affairs 2018]{UN2018}
United~Nations, D. o.~E. and Affairs, S. (2018).
\newblock {\em World Urbanization Prospects: The 2018 Revision, Highlights}.
\newblock United Nations.

\bibitem[Vaquero and Rodero-Merino 2014]{Vaquero:2014:FYW:2677046.2677052}
Vaquero, L.~M. and Rodero-Merino, L. (2014).
\newblock Finding your way in the fog: Towards a comprehensive definition of
  fog computing.
\newblock {\em SIGCOMM Comput. Commun. Rev.}, 44(5):27--32.

\bibitem[Wang et~al. 2017]{wang2017road}
Wang, H., Wen, H., Yi, F., Zhu, H., and Sun, L. (2017).
\newblock Road traffic anomaly detection via collaborative path inference from
  gps snippets.
\newblock {\em Sensors}, 17(3):550.

\bibitem[Wang et~al. 2018]{wang2018city}
Wang, X., Ning, Z., Hu, X., Ngai, E. C.-H., Wang, L., Hu, B., and Kwok, R.~Y.
  (2018).
\newblock A city-wide real-time traffic management system: Enabling
  crowdsensing in social internet of vehicles.
\newblock {\em IEEE Communications Magazine}, 56(9):19--25.

\bibitem[Wen et~al. 2017]{Wen2017}
Wen, Z., Yang, R., Garraghan, P., Lin, T., Xu, J., and Rovatsos, M. (2017).
\newblock Fog orchestration for internet of things services.
\newblock {\em IEEE Internet Computing}, 21(2):16--24.

\bibitem[Xu and Li 2013]{Xu2013}
Xu, H. and Li, B. (2013).
\newblock Joint request mapping and response routing for geo-distributed cloud
  services.
\newblock In {\em 2013 Proceedings IEEE INFOCOM}, pages 854--862.

\bibitem[Xu et~al. 2014]{IoT:2014}
Xu, L.~D., He, W., and Li, S. (2014).
\newblock Internet of things in industries: A survey.
\newblock {\em IEEE Transactions on Industrial Informatics}, 10(4):2233--2243.

\bibitem[Yabe et~al. 2016]{yabe2016framework}
Yabe, T., Tsubouchi, K., Sudo, A., and Sekimoto, Y. (2016).
\newblock A framework for evacuation hotspot detection after large scale
  disasters using location data from smartphones: case study of kumamoto
  earthquake.
\newblock In {\em Proceedings of the 24th ACM SIGSPATIAL International
  Conference on Advances in Geographic Information Systems}, page~44. ACM.

\bibitem[Yang et~al. 2017]{Yang2017}
Yang, C., Huang, Q., Li, Z., Liu, K., and Hu, F. (2017).
\newblock Big data and cloud computing: innovation opportunities and
  challenges.
\newblock {\em International Journal of Digital Earth}, 10(1):13--53.

\bibitem[Yi et~al. 2015]{yi2015fog}
Yi, S., Hao, Z., Qin, Z., and Li, Q. (2015).
\newblock Fog computing: Platform and applications.
\newblock In {\em 2015 Third IEEE Workshop on Hot Topics in Web Systems and
  Technologies (HotWeb)}, pages 73--78. IEEE.

\bibitem[Zhang et~al. 2017]{zhang2017mobile}
Zhang, K., Mao, Y., Leng, S., He, Y., and Zhang, Y. (2017).
\newblock Mobile-edge computing for vehicular networks: A promising network
  paradigm with predictive off-loading.
\newblock {\em IEEE Vehicular Technology Magazine}, 12(2):36--44.

\bibitem[Zheng et~al. 2015]{zheng2015heterogeneous}
Zheng, K., Zheng, Q., Chatzimisios, P., Xiang, W., and Zhou, Y. (2015).
\newblock Heterogeneous vehicular networking: a survey on architecture,
  challenges, and solutions.
\newblock {\em IEEE communications surveys \& tutorials}, 17(4):2377--2396.

\bibitem[Zheng 2011]{Zheng2011}
Zheng, Y. (2011).
\newblock {\em Location-Based Social Networks: Users}, pages 243--276.
\newblock Springer New York, New York, NY.

\bibitem[Zheng 2012]{Zheng2012}
Zheng, Y. (2012).
\newblock {Tutorial on Location-Based Social Networks}.
\newblock In {\em Proc.\ of WWW'12}, Lyon, France.

\bibitem[{Zheng} 2015]{zheng2015}
{Zheng}, Y. (2015).
\newblock Methodologies for cross-domain data fusion: An overview.
\newblock {\em IEEE Transactions on Big Data}, 1(1):16--34.

\bibitem[Zheng et~al. 2014]{zheng2014urban}
Zheng, Y., Capra, L., Wolfson, O., and Yang, H. (2014).
\newblock Urban computing: concepts, methodologies, and applications.
\newblock {\em ACM Transactions on Intelligent Systems and Technology (TIST)},
  5(3):38.

\bibitem[Zhu et~al. 2018]{zhu2018big}
Zhu, L., Yu, F.~R., Wang, Y., Ning, B., and Tang, T. (2018).
\newblock Big data analytics in intelligent transportation systems: a survey.
\newblock {\em IEEE Transactions on Intelligent Transportation Systems},
  20(1):1--16.

\bibitem[Zografos and Androutsopoulos 2008]{zografos2008algorithms}
Zografos, K.~G. and Androutsopoulos, K.~N. (2008).
\newblock Algorithms for itinerary planning in multimodal transportation
  networks.
\newblock {\em IEEE Transactions on Intelligent Transportation Systems},
  9(1):175--184.

\end{thebibliography}
   
\end{document}